\documentclass[twocolumn,dvipsnames]{aastex631}

\usepackage{apjfonts}
\DeclareSymbolFont{cmletters}{OML}{cmm}{m}{it}
\DeclareMathSymbol{v}{\mathalpha}{cmletters}{"76} 
\usepackage{hyperref}
\usepackage{booktabs}
\usepackage{amssymb}
\usepackage{amsmath}

\usepackage{xspace}

\newcommand{\chandra}{{\it Chandra}\xspace}

\newcommand{\eqb}{\begin{eqnarray}}
\newcommand{\eqe}{\end{eqnarray}}

\newcommand{\mdot}{\ensuremath{\dot M}\xspace}

\newcommand{\mdotb}{\ensuremath{\dot M_{\rm B}}\xspace}
\newcommand{\rh}{\ensuremath{R_{\rm H}}\xspace}
\newcommand{\rg}{\ensuremath{R_{\rm g}}\xspace}
\newcommand{\rb}{\ensuremath{R_{\rm B}}\xspace}
\newcommand{\rcirc}{\ensuremath{R_{\rm circ}}\xspace}
\newcommand{\BH}{{{\rm BH}}\xspace}
\newcommand{\SMBH}{{{\rm SMBH}}\xspace}
\newcommand{\MBH}{{M_{\rm BH}}\xspace}
\newcommand{\phibh}{{\phi_{\rm BH}}\xspace}
\newcommand{\der}{{\rm d}}
\newcommand{\p}{{\partial}}
\newcommand{\K}{\ensuremath{{\rm k}}}

\newcommand{\stateone}{MAD\xspace}
\newcommand{\statetwo}{BAD\xspace}
\newcommand{\statethree}{RAD\xspace}

\usepackage{ulem}

\usepackage{hyperref}

\shorttitle{Global Jet Destruction}
\shortauthors{Lalakos et al.}

\graphicspath{{./}{figures/}}
\usepackage{graphicx}

\usepackage[sort&compress]{natbib}
\defcitealias{bromberg2016}{BT16}

\defcitealias{Lalakos2022}{L22}
\defcitealias{blanford1977}{BZ}
\defcitealias{ressler2021magnetically}{R21}
\defcitealias{2023ApJ...946L..42K}{K23}

\begin{document}

\title{Jets with a Twist: Emergence of FR0 Jets in 3D GRMHD Simulation of Zero Angular Momentum Black Hole Accretion}

\author[0000-0002-6883-6520]{Aretaios Lalakos}
\email{lalakos@u.northwestern.edu}
\affiliation{Center for Interdisciplinary Exploration \& Research in Astrophysics (CIERA), Physics \& Astronomy, Northwestern University, Evanston, IL 60202, USA}

\author[0000-0002-9182-2047]{Alexander Tchekhovskoy}
\affiliation{Center for Interdisciplinary Exploration \& Research in Astrophysics (CIERA), Physics \& Astronomy, Northwestern University, Evanston, IL 60202, USA}

\author[0000-0003-4271-3941]{Omer Bromberg}
\affiliation{The Raymond and Beverly Sackler School of Physics and Astronomy, Tel Aviv University, Tel Aviv 69978, Israel}

\author[0000-0003-3115-2456]{Ore Gottlieb}
\affiliation{Center for Interdisciplinary Exploration \& Research in Astrophysics (CIERA), Physics \& Astronomy, Northwestern University, Evanston, IL 60202, USA}
\affiliation{Center for Computational Astrophysics, Flatiron Institute, New York, NY 10010, USA}
\affiliation{Department of Physics and Columbia Astrophysics Laboratory, Columbia University, Pupin Hall, New York, NY 10027, USA}

\author[0000-0003-2982-0005]{Jonatan Jacquemin-Ide}
\affiliation{Center for Interdisciplinary Exploration \& Research in Astrophysics (CIERA), Physics \& Astronomy, Northwestern University, Evanston, IL 60202, USA}

\author[0000-0003-4475-9345]{Matthew Liska}
\affiliation{Center for Relativistic Astrophysics, Georgia Institute of Technology, Howey Physics Bldg, 837 State St NW, Atlanta, GA 30332, USA}
\affiliation{Institute for Theory and Computation, Harvard University, 60 Garden Street, Cambridge, MA 02138, USA}

\author[0000-0001-9826-1759]{Haocheng Zhang}
\affiliation{Center for Space Sciences and Technology, University of Maryland Baltimore County, Baltimore, MD 21250}
\affiliation{NASA Goddard Space Flight Center, Greenbelt, MD 20771, USA}

\begin{abstract}
Spinning supermassive black holes (BHs) in active galactic nuclei (AGN) magnetically launch relativistic collimated outflows, or jets. Without angular momentum supply, such jets are thought to perish within $3$ orders of magnitude in distance from the BH, well before reaching kpc-scales. We study the survival of such jets at the largest scale separation to date, via 3D general relativistic magnetohydrodynamic simulations of rapidly spinning BHs immersed into uniform zero-angular-momentum gas threaded by weak vertical magnetic field. We place the gas outside the BH sphere of influence, or the Bondi radius, chosen much larger than the BH gravitational radius, $R_\text{B}=10^3R_\text{g}$. The BH develops dynamically-important large-scale magnetic fields, forms a magnetically-arrested disk (MAD), and launches relativistic jets that propagate well outside $R_\text{B}$ and suppress BH accretion to $1.5$\% of the Bondi rate, \mdotb.  Thus, low-angular-momentum accretion in the MAD state can form large-scale jets in Fanaroff-Riley (FR) type I and II galaxies. Subsequently, the disk shrinks and exits the MAD state: barely a disk (\statetwo), it rapidly precesses, whips the jets around, globally destroys them, and lets $5-10$\% of \mdotb reach the BH. Thereafter, the disk starts rocking back and forth by angles $90-180^\circ$: the rocking accretion disk (\statethree) launches weak intermittent jets that spread their energy over a large area and suppress BH accretion to $\lesssim2$\% \mdotb. Because BAD and RAD states tangle up the jets and destroy them well inside $R_\text{B}$, they are promising candidates for the more abundant, but less luminous, class of FR0 galaxies.
\end{abstract}


\keywords{High energy astrophysics (739); Active galactic nuclei (16); Black hole
physics (159); Jets (870); Magnetohydrodynamical simulations (1966); General relativity (641)}


\section{Introduction} \label{sec:intro}
Relativistic collimated outflows, known as jets, are prevalent across many astrophysical systems of vastly different scales. The largest and most energetic ones emanate from the galaxy centers that harbor supermassive black holes (\SMBH{}s). Active galactic nuclei (AGN), whose central \SMBH{}s consume gas, release energy that couples to the galactic environment in a process called AGN feedback. The general consensus is that the feedback comes in two flavors: (i)  Radio or kinetic mode, which occurs at lower accretion rates ($L\ll 0.01 L_{\rm Edd}/c^2$, where $L_{\rm Edd}$ is the Eddington luminosity). In this mode, powerful radio jets dominate the feedback (see \citealt{fabian2012,morganti2017many}, for reviews). (ii) Radiative or quasar mode, which takes place at high accretion rates in luminous AGN ($L\gtrsim 0.01 L_{\rm Edd}/c^2$). Here, the radiation coupled with the surrounding gas and dust drives powerful outflows at wide angles into the galactic environment. About 1 in 10 luminous AGN can produce powerful radio jets \citep{ssl07}. In both cases, (i) and (ii), the radio jets can propagate into the interstellar medium (ISM) and intracluster medium (ICM), displace the gas, and inflate X-ray cavities of up to several Mpc in size (\citealt{2022A&A...660A...2Ow}; see \citealt{mcnamara2007heating, mcnamara2012mechanical}, for reviews). Whereas these cavities appear empty in X-ray images, they are filled with relativistically-hot magnetized plasma that emits copious synchrotron emission in the radio band. 
The cavities result from AGN feedback that can offset the runaway cooling of the ICM via shocks, acoustic waves, and/or turbulent heating and thus can regulate star formation \citep{2016MNRAS.458.2902Z,martizzi2019simulations,2020ApJ...889L...1L}. 
However, how \SMBH{}s feed on the gas, power the jets, and exert feedback on their environment remains poorly understood. 

The Fanaroff-Riley (FR) classification system \citep{fr74} is widely used to categorize radio galaxies into two main types: FRIs and FRIIs, determined by the morphology of their radio emission. FRIs are bright near the core and grow gradually fainter with increasing distance. FRIIs are brighter near the edge of the radio lobes and grow fainter towards the core. Recently, a new classification has been established, the FR0s that morphologically lack radio emission at scales $r\gtrsim 1$ kpc \citep{ghisellini2011extragalactic}. They are more abundant than FRIs and FRIIs \citep{stecker2019extragalactic}, hinting that they might be the dominant population in the local universe (\citealt{refId0}; see \citealt{baldi2023nature} for review). Additionally, the Fermi observatory recently detected bright $\gamma$-rays from $3$ out of $\gtrsim100$ FR0 galaxies, with $2$ being an order of magnitude brighter in radio than the rest \citep{Paliya_2021}. This, along with their abundance, suggests that FR0s can significantly contribute to the isotropic $\gamma-$ray background \citep{stecker2019extragalactic}, the diffuse neutrino background \citep{10.1093/mnras/sty251,ackermann2022high}, as well as ultra-high-energy cosmic rays \citep{merten2021scrutinizing,lundquist2021extrapolating}.
It is not clear yet whether FR0s are part of the evolutionary stage of FRI/IIs. The SMBH mass and host environment appear similar to the FRIs, however, the jet size is smaller, $r\lesssim 1$kpc, usually unresolved, and moves slower $\gamma \lesssim 2$ \citep{balmaverde2006host,refId0}. This might suggest that either the BH spin or the magnetic field strength are different enough that can affect the jet power and velocity \citep{refId0}.

Spinning \BH{}s surrounded by whirlpools of hot magnetized plasma, or accretion disks, can produce relativistic jets via the \citet[][\citetalias{blanford1977} hereafter]{blanford1977} process. The \citetalias{blanford1977} process capitalizes on the fact that as the hot plasma in the disk accretes on a spinning \BH, it brings with it poloidal (pointing in the $R$- and $z$-directions) magnetic flux. \BH rotation drags the inertial frames via the \citet{1918PhyZ...19..156L} effect, which twists the poloidal flux in the vicinity of the \BH, causing it to build up a strong toroidal component, which in turn powers collimated Poynting-flux dominated outflows -- the twin polar jets  (see \citealt{2015ASSL..414...45T}). The rotation of the disk launches disk winds (via a \citealt{bp82} like process) that collimate the jets into small opening angles and enable the jets to accelerate \citep{komissarov_magnetic_2008,tch09}.
This suggests that \emph{rotation} is an important ingredient for accretion, jets and outflows, and their feedback on the environment. 
In this work, we aim to reveal the importance of ambient gas rotation on the jets and their feedback.

To understand how the jets escape out of the \BH sphere of influence, one must follow them from their formation at the \BH event horizon to their interaction with the ISM or ICM. For this, one needs
to model the \BH feeding by following the gas infall from the ISM to the \BH event horizon. These are extremely arduous tasks due to the enormous scale separation of the problem. The gas originates at the edge of the BH sphere of influence, the Bondi radius \citep{bon52}, 
\begin{equation}
\rb = \cfrac{G\MBH}{c^2_{\infty}} \sim 20\ {\rm pc} \times \cfrac{\MBH}{10^9M_\odot}\times \left (\cfrac{T_\infty}{1 \text{keV}} \right )^{-1},
\label{eq:rb}
\end{equation}
which is significantly larger than the \BH gravitational radius, $\rg = G\MBH/c^2 = 5\times10^{-5}\ {\rm pc}\times(\MBH/10^9M_\odot)$.  Here, $\MBH$ is the \SMBH mass and $c_\infty$ is the ISM sound speed. For a typical ISM temperature, $T_{\rm \infty}\sim 1$~keV, the scale separation between \rb and \rg reaches 6 orders of magnitude \citep{russell2015, russell2018}. Adding to the challenge is that the ICM and dark matter halos reach scales of $\sim 100$~kpc, marking the scale separation of the full problem $\sim 9-10$ orders of magnitude. Dedicated galaxy simulations are capable of following gas flows from the dark matter halo down to $\sim 1$~kpc scales \citep{alcazar2015,alcazar2017},  and state-of-the-art hyper-refined Lagrangian simulations can even reach sub-pc scales \citep{2021ApJ...917...53A}. However, to bridge the ``last mile'' of the scale separation and connect the smallest scales in galaxy simulations to the event horizon, general relativistic magnetohydrodynamic (GRMHD) simulations are essential. 

A popular approach for bridging the scale separation is the Bondi model \citep{bon52, shapiro_black_holes_1986}, which approximates \BH accretion as a spherically symmetric hydrodynamic flow. Although an elegant and simple approximation, it does not allow for jets.
In this work, we add the minimum ingredients to the Bondi model that enable the study of BH-powered jet formation and propagation: we retain the zero angular momentum of the ambient gas but add \BH rotation and ambient vertical magnetic field. We also consider the system at sufficiently high resolution and in full 3D to resolve the jet propagation and allow for the development of non-axisymmetric instabilities in the jets.

GRMHD simulations found that magnetized accretion can accumulate large-scale poloidal magnetic fields on the \BH to the point that the fields become dynamically important, i.e., able to obstruct gas infall, at which point the system enters the magnetically arrested disk state \citep[MAD,][]{1974Ap&SS..28...45B, 1976Ap&SS..42..401B, nia03, igu03, 2008ApJ...677..317I} that can launch jets with energy efficiencies exceeding $100$\%, i.e., whose power exceeds the accretion power \citep{tchekhovskoy2011efficient}. In this state, the accretion continuously brings in the magnetic flux to the BH, thereby flooding both the BH and inner disk with the vertical magnetic flux. At the same time, the dynamically-important magnetic flux periodically erupts from the BH, rips through the disk, and escapes. In the presence of sufficient angular momentum, the inward advection of the magnetic flux wins, and the BH is always flooded with the magnetic flux \citep{2012MNRAS.423L..55T}.

In contrast, zero angular momentum accretion was found to not be conducive to powerful, stable jet production. Analytic models suggested that zero angular momentum accretion produces weak jets  \citep[with energy efficiency $\lesssim 1$\%,][]{2012MNRAS.421L..24D}. 
%
%
\citet[hereafter \citetalias{2023ApJ...946L..42K}]{2023ApJ...946L..42K} used GRMHD simulations to model the accretion of spherically-symmetric magnetized gas on a rapidly spinning \BH and found that zero-angular accretion could not remain in a MAD state for a prolonged period of time. The jet efficiency only transiently surpassed $100\%$ due to the BH eventually losing the large-scale BH magnetic flux powering the jets \citep[see also][for a related phenomenon in collapsars]{Gottlieb2022black} and entering the SANE state \citep[``standard and normal evolution'',][]{2012MNRAS.426.3241N}, in which the large-scale magnetic flux is sub-dominant and the jets are weaker or non-existent.
%
%
\citet[hereafter \citetalias{ressler2021magnetically}]{ressler2021magnetically} performed 3D GRMHD simulations of spherically symmetric gas accretion onto a rapidly-spinning BH ($a=0.9375$), with a Bondi-to-event horizon scale separation of $\rb/\rg = 100$.\footnote{Note that in comparison to our eq.~\eqref{eq:rb}, \citetalias{ressler2021magnetically} use an extra factor of $2$ in their Bondi radius definition: $\rb = 2G\MBH/c_{\rm \infty}^2=200 \rg$.} They varied the angle between the ambient magnetic field and BH spin directions and found that outflow efficiencies, in most cases, did not exceed $100\%$: the accretion flow had a hard time reaching and staying MAD. Based on the results of their simulations, \citetalias{ressler2021magnetically} predicted that for $\rb/\rg \gtrsim 800$ all jets powered by low angular momentum accretion will end up falling victim to the kink instability inside the Bondi radius, as is the case for realistic systems, $\rb/\rg \gtrsim 10^{5-6}$ (e.g., SgrA* or M87). 

Here, we evaluate the ability of low angular momentum accretion to power large-scale jets for an order of magnitude larger scale separation than has been possible until now.
\citet[hereafter \citetalias{Lalakos2022}]{Lalakos2022} simulated 3D GRMHD accretion of rotating ISM for the Bondi-to-event horizon ratio of $\rb/\rg=10^3$. Here, we follow the \citetalias{Lalakos2022} approach, but consider zero angular momentum accretion. As in \citetalias{Lalakos2022}, we include a rapidly spinning \BH, $a=0.94$, and large-scale vertical magnetic field, which is aligned with the BH spin vector.
To study the self-consistent jet formation and propagation to distances well outside the Bondi radius, we use adaptive mesh refinement (AMR) to ensure sufficient resolution of the tightly collimated jets. 

More generally, we aim to understand the basic ingredients necessary for the formation of stable jets or, conversely, what it takes to ``break'' them. Namely, we aim to answer questions such as: Is gas angular momentum (and the formation of an accretion disk) needed to maintain jet stability?  Is there a critical power above which the jets manage to escape the BH sphere of influence? How does the ambient medium trigger jet instability? What are the observational signatures of unstable jets? 

Throughout the paper, we work in units of $G=M=c=1$. For conciseness, we sometimes measure the time in the unit of $1000 \rg/c$, which we denote as $1$k: e.g., $7.5\times 10^4 \rg/c \equiv 75$k. In Sec.~\ref{sec:Numerical setup}, we describe our simulation setup and choice of the initial physical parameters. In Sec.~\ref{sec:infall}, we present the properties of the accretion flow and jets as measured near the \BH. In Sec.~\ref{sec:kink}, we show how the jets can escape out of the Bondi sphere and discuss their stability. In Sec.~\ref{sec:epilogue} we show how the changes in the behavior of accretion flow angular momentum can lead to the destruction of the jets. In Sec.~\ref{sec:ext-vs-int-kink}, we compare internal and external kink instabilities. Finally, in Sec.~\ref{sec:conclusions}, we summarize and discuss our results.
		
	

\section{Numerical method and setup} \label{sec:Numerical setup}

We carry out our simulations using the general relativistic magnetohydrodynamic (GRMHD) code {\sc h-amr} that includes advanced features such as graphics processing unit (GPU) acceleration, AMR, and local adaptive timestepping (\citealt{liska2022h}). We initialize the simulation by placing a \BH inside a uniform static ambient medium. We consider a rapidly spinning \BH, with dimensionless spin $a=0.94$, to favor jet launching. 
To allow the system to evolve naturally, we avoid prescribing the conditions inside \rb: we carve out an empty cavity inside the Bondi sphere ($r<\rb$) and place the ambient gas of uniform rest-mass density, $\rho =\rho_{\infty}$, outside the sphere ($r\ge\rb$). 

We choose the ambient sound speed, $c_\infty$, to achieve the desired scale separation, $\rb=10^3\rg$ (eq.~\ref{eq:rb}). This is similar to \citetalias{Lalakos2022}, but here we set the ambient medium angular momentum to zero. 
In this work, we focus on low-luminosity \BH accretion systems. Because of this, we do not include any radiation effects (e.g., radiative cooling), which are important at higher mass accretion rates. Thus, our simulations are non-radiative and scale-free: the simulation results trivially rescale to any value of $\rho_{\infty}$. As appropriate for AGN environments at the Bondi radius, we adopt a monatomic non-relativistic ($\Gamma=5/3$) gas with an ideal gas equation of state, $p_{\rm g}=(\Gamma-1)u_{\rm g}$,  where $p_{\rm g}$ and $u_{\rm g}$ are the pressure and internal energy density of the gas. 

Outside the Bondi radius, we include a large-scale vertical magnetic field in the direction parallel to the \BH spin. Asymptotically far away, we set $\vec{B}=B_0\vec a$. Closer to \rb, we deform the field such that the radial component of the magnetic field,  $B^r$, smoothly vanishes towards the edge of the cavity and no field enters the cavity, $r\le\rb$. To achieve this, we adopt the magnetic vector potential, $A_{\varphi}\propto(r^2-\rb^2) \sin^2\theta$.  We normalize the strength of the magnetic field such that the thermal to magnetic pressure ratio is $\beta=p_{\rm g}/p_{\rm m}=100$ asymptotically far away (this also ensures that $\beta\ge100$ everywhere). Here, $p_{\rm m}=b^2/8\pi$ is the magnetic pressure, where $b^2 = b^\mu b_\mu$, and $b^\mu$ is the comoving contravariant magnetic field 4-vector (defined in Appendix~\ref{App:comoving_fields})
In order to break axisymmetry and provide the seeds for the growth of the magneto-rotational instability (MRI, \citealt{bal91}), we include random pressure perturbations at a $2\%$ level in the initial conditions. 

We note that the cavity is, technically, not entirely empty because GRMHD codes cannot handle vacuum. To prevent $\rho$ and $u_g$  from becoming extremely low or negative in highly magnetized regions (i.e., the jet launching regions), our numerical scheme adopts the following density floors. If at any point in the simulation the densities drop below the floor values, $\rho < \rho_{\rm fl} = \max[b^2/(60\pi),10^{-7} r^{-2},10^{-20}]$ and/or $u_{\rm g} < u_{\rm {g, fl}} = \max[b^2/(3000\pi),10^{-9} r^{-2 \Gamma},  10^{-20}]$, then we add mass and/or internal energy, respectively, in the drift frame until the floor values are reached \citep[see Appendix B3 of][]{2017MNRAS.467.3604R}.


We construct our grid in spherical polar coordinates, $r$, $\theta$, and $\varphi$.  The radial grid is uniform in $\log r$, and $r$ spans $0.83\rh\le r \le 10^6\rg$. There are 6 cells inside the event horizon, $\rh = \rg(1+\sqrt{1-a^2})$, and the radius of the outer boundary is larger than the light travel distance in a simulation duration: these ensure that both the inner and outer radial boundaries are causally disconnected from and cannot influence the solution. The polar and azimuthal grids are uniform in the $\theta$- and $\varphi$-directions, and span $0 \le \theta \le \pi$ and $0\le\varphi\le2\pi$, respectively. We use outflow, transmissive, and periodic boundary conditions in the radial, polar, and azimuthal directions, respectively \citep{liska2022h}.  The base-grid resolution is $N_r\times N_{\theta} \times N_{\varphi}=448\times96\times192$ cells in the \hbox{$r$-,} \hbox{$\theta$-,} and \hbox{$\varphi$-}directions. At $r\ge6.5\rg$, we activate 1 level of static mesh refinement (SMR): this doubles the resolution in each dimension, leading to an increased effective resolution of $896\times192\times384$. On top of the SMR, we also include 2 additional adaptive mesh refinement (AMR) levels that we dynamically activate during the run to ensure sufficient resolution to resolve the collimated jets and cocoons at large radii as we describe in Appendix~\ref{app:amr-criterion}. The maximum effective resolution in the jets can therefore reach $3,584\times768\times1,536$ cells.

In order to avoid the polar singularity interfering with highly magnetized jetted regions, we tilt the entire BH-gas system by $90^{\rm \circ}$ relative to our computational grid: as a result, the \BH rotational axis is perpendicular to the polar singularity of our computational grid. However, for the sake of narrative simplicity, below we present our results as if we did not perform the tilt: for presentation purposes, we direct the $z$-axis along the \BH spin vector and count off the polar angle, $\theta$, from the \BH spin direction.


\section{MAD Prologue} \label{sec:infall}
The simulation starts with the constant-density ambient gas at rest, located outside of the empty cavity, $r\ge\rb$ (Sec.~\ref{sec:Numerical setup}). The pressure gradient at the edge of the cavity, along with the gravitational attraction from the \BH, push the gas inward.
To study the mass flow in our simulation, we define the rest-mass (RM) component of the energy flux density, 
\begin{equation}
    f_{\rm RM} = \rho c^2 u^r,
\end{equation}
whose angular integral gives us the net energy rate, or power,
\begin{equation}
    \dot{E}_{\rm RM} = \iint  f_{\rm RM} \der A,
    \label{eqn:RM_fl}
\end{equation}
where $\der A=\sqrt{-g}\der\theta \der\varphi$ is the differential surface element, $g= \left|g_{\mu \nu}\right|$ is the determinant of the metric, and $u^\mu$ is the coordinate-frame contravariant proper velocity vector. In a steady state, $\dot{E}_{\rm RM}$ is conserved and independent of radius. We evaluate \BH mass accretion rate as the negative of the rest-mass power, $\mdot \equiv -\dot{E}_{\rm RM}(r=8\rg)/c^2$,  which we measure at $r = 8\rg$ to avoid potential contamination by the density floors near the event horizon. 

\begin{figure*}[!t]
\centering
\includegraphics[width=0.9\textwidth]{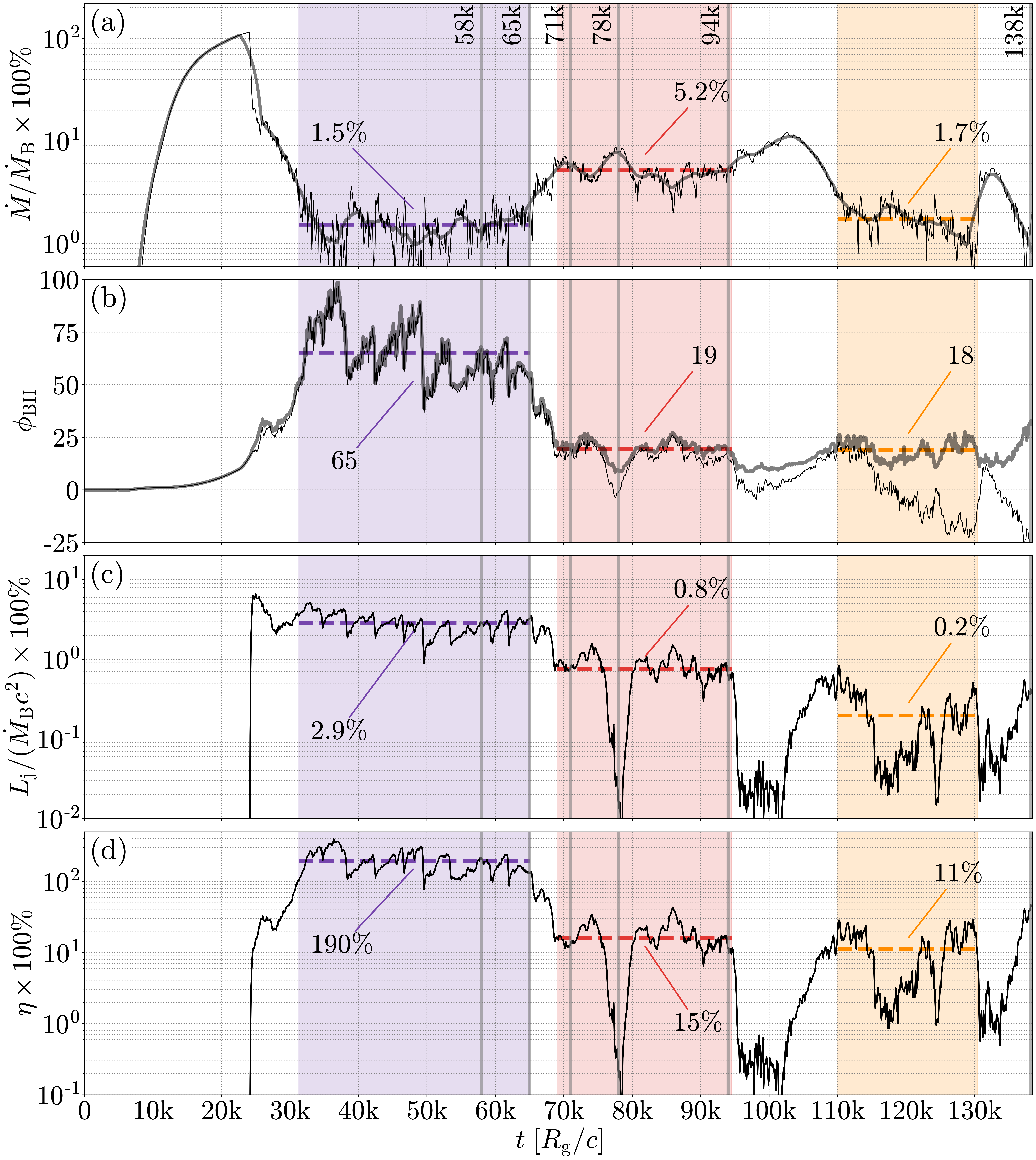}
\caption{
In the absence of rotation and a strong magnetic field, the instantaneous \BH mass accretion rate can transiently reach the Bondi value, $\mdot \simeq \mdotb$ (thin line in {\bf panel~a}). However, once the magnetic flux accumulates on the \BH event horizon and its normalized absolute value exceeds $\phibh \approx 15$ at $t=24$k (thick line in {\bf panel~b}), the first jets -- of normalized power $L_{\rm j}/\mdotb c^2 \simeq 5\%$ ({\bf panel~c}) and efficiency $\eta \simeq 10\%$ ({\bf panel~d}) -- emerge and suppress the instantaneous \BH accretion rate to $\mdot/\mdotb \simeq 20$\% (thin line in {\bf panel~a}; we normalize $\phibh$ and $\eta$ to $\langle\mdot\rangle_\tau$, which is $\mdot$ smoothed over $\tau =3$k and shown with the thick line in {\bf panel~a}). Following this initial transient phase, the accumulating magnetic flux reaches $\phibh\gtrsim50$ (thick line in {\bf panel~b}) and leads to a magnetically arrested disk state (MAD, purple-shaded region, $31\K\le t\le65$k). In a MAD, jets of high power, $\left<L_{\rm j}\right>/\mdotb c^2 \simeq 2.9\%$ ({\bf panel~c}), and efficiency, $\left< \eta \right> \simeq 190\%$ ({\bf panel~d}), severely suppress \BH accretion to $\langle\mdot\rangle/\mdotb \simeq 1.5$\% ({\bf panel~a}, Sec.~\ref{sec:infall}). At $t\gtrsim 65$k, the system exits the MAD state, the jets become weaker, $\left <\eta \right> \lesssim 15\%$, and the mass accretion rate higher, $\left <\mdot \right>/\mdotb \lesssim 5$\% (red-shaded region, $69\K\le t\le 95$k). The jets sometimes completely shutoff and fall apart (e.g., at $t=78$k and $96-102$k; Sec.~\ref{sec:kink}) when the normalized Northern hemisphere \BH magnetic flux vanishes, $\phi_{\rm N}=0$ (thin line in {\bf panel~b}), $\eta$ drops (to $\lesssim0.1$\%, {\bf panel~d}), and the reduced jet feedback allows higher $\mdot/\mdotb\lesssim10$\% ({\bf panel~a}). However, even such weaker jets can suppress \mdot similar to the MAD state, $\left<\mdot\right>/\mdotb \simeq 1.7$\%, for extended periods of time if they continuously reorient (orange-shade region $110\K\le t \le130$k; Sec.~\ref{sec:epilogue} and Fig.~\ref{fig:angular_momentum}). In all panels, horizontal dashed lines show time average values, which are also labeled with call-outs.
}
\label{fig:panel_fig}
\end{figure*}

Figure~\ref{fig:panel_fig}(a) shows the time-dependence of $\mdot$, which peaks at approximately the analytic Bondi prediction,
\begin{equation}
\label{eqn:analytics_Bondi}
\mdotb = 4 \pi \lambda_{\rm s}(GM_{\rm \BH})^2 \cfrac{\rho_{\infty}}{ c_{\infty}^{3}}\,,
\end{equation}
at about a free-fall time, $t_{\rm ff} = 2^{-1/2} (\rb/\rg)^{3/2} \rg/c \simeq 2.2\times 10^4 \rg/c = 22$k, after the beginning of the simulation. Here, for our choice, $\Gamma=5/3$, we have $\lambda_{\rm s}=1/4$  \citep{Shapiro1976, di2003accretion}. 
That $\mdot$ reaches the simple analytic Bondi prediction, $\mdotb$, is not entirely surprising because, in the absence of any external angular momentum supply, the gas never encounters a centrifugal barrier, which would inhibit the accretion relative to the Bondi expectation.

The infalling gas drags inward the large-scale vertical magnetic flux, and some of it makes it all the way to the event horizon, resulting in the increase of the absolute \BH magnetic flux, 
\begin{equation}
\label{eq:PhiBH}
\Phi_{\rm BH} = 0.5 \iint \left | B^r \right | \der A. 
\end{equation}
Here, the integral is over the entire event horizon of the \BH, and the factor of $0.5$ converts it to a single hemisphere. Figure~\ref{fig:panel_fig}(b) shows the time dependence of normalized absolute magnetic flux,
\begin{equation}
  \label{eq:phibh}
  \phibh=\frac{\Phi_{\rm BH}}{\sqrt{\left <\dot{M} \right >_\tau\rg^2 c}},
\end{equation}
which measures the strength of the magnetic flux relative to the onslaught of the infalling gas; here, to clarify the units, we explicitly include the length and velocity scaling factors, and $\left <\dot{M} \right >_\tau$ is the rolling average of the mass accretion rate over the time interval of $\tau = 3$k. This time interval is sufficiently long to average over the strong \mdot oscillations in the MAD state. We also consider a signed magnetic flux through the Northern hemisphere, 
\begin{equation}
\label{eq:PhiN}
\Phi_{\rm N} = \int_0^{\pi/2}\der\theta\int_0^{2\pi}\der\varphi B^r \sqrt{-g}, 
\end{equation}
as well as its normalized form, $\phi_{\rm N}$, defined analogous to eq.~\eqref{eq:phibh}. As the infalling gas continuously drags more of the ambient magnetic flux inwards, the \BH magnetic flux grows in strength and $\phibh$ steadily increases. 

Similar to eq.~\eqref{eqn:RM_fl}, we compute the various components of the energy flux using the stress-energy tensor:
\begin{equation}
    T^\kappa_\lambda = \left( \rho c^2+ u_{\rm g} + p_{\rm g} + \frac{b^2}{4\pi} \right ) u^\kappa u_\lambda + \left( p_{\rm g} + \frac{b^2}{8\pi} \right) \delta^\kappa_\lambda - \frac{b^\kappa b_\lambda}{4\pi}
    \label{eqn:TOT_fl}
\end{equation}
We are interested in the radial energy flux and set $\kappa=r$ and $\lambda=t$. Thus evaluated, eq.~\eqref{eqn:TOT_fl}  gives us the total radial energy flux, $f_{\rm TOT}$. The electromagnetic (EM), thermal (TH), and kinetic energy (KE) flux density components of the total energy flux, respectively, are:
\begin{align}
    f_{\rm EM} &= - \left(b^2 u^r u_t -b^r b_t\right)/4\pi, 
    \label{eqn:EM_fl}\\
    f_{\rm TH} &= - \left( u_{\rm g} + p_{\rm g} \right ) u^r u_t,
    \label{eqn:TH_fl}\\
    f_{\rm KE} &= - \rho c^2 u^r u_t,
    \label{eqn:KE_fl}
\end{align}
where
$u^r$ is the radial contravariant 4-velocity component and $u_{t}$ is the temporal covariant velocity component, which is a conserved quantity for point masses. We also define the components of power (integral versions of the flux densities), eqs.~\eqref{eqn:EM_fl}--\eqref{eqn:KE_fl}, following eq.~\eqref{eqn:RM_fl}: 
\begin{equation}
\dot{E}_\# \equiv \iint f_\# \der A,
\label{eq:Fint}
\end{equation}
where $\# =$ EM, TH, KE, or RM. The integral version of eq.~\eqref{eqn:EM_fl} gives us the jet power, $L_{\rm j}=\dot{E}_{\rm EM}(r=8\rg)$, which we measure through a sphere of radius $r = 8\rg$. Note that  $L_{\rm j}$ includes the power of both jets. Figure~\ref{fig:panel_fig}(c) shows that $L_{\rm j}$ abruptly increases at $t\simeq 24$k once $\phibh$ exceeds a critical value, $\phibh\simeq 15$, and the jets form. 

Figure~\ref{fig:panel_fig}(d) shows that the outflow energy efficiency, $\eta \equiv L_{\rm j}/\langle \dot Mc^2\rangle_\tau$, which is jet power measured in units of \BH accretion power: at $t\simeq 24$k, it is not very high yet, but it is still significant, $\eta = 0.1 \equiv 10$\%, where from now on we will express all fractions in terms of percent. Although at this time the jets have not yet reached the maximum efficiency, they already exert significant feedback on the accretion flow and suppress \BH accretion rate relative to its peak by nearly an order of magnitude, $\mdot \sim 0.2\mdotb$, via injecting the energy into the accretion flow and partially unbinding it. This marks the end of the initial transient phase, during which the system settles into a quasi-steady state. 

Figure~\ref{fig:panel_fig}(b) shows that the dimensionless \BH magnetic flux grows until it saturates around $\phibh\gtrsim 50$ at $t\gtrsim 31$k. At this time, the \BH magnetic flux becomes dynamically important, i.e., the magnetic pressure can withstand the onslaught of the total momentum flux of the infalling gas. This signals the formation of a MAD \citep{tchekhovskoy2011efficient}. The jets produced during the MAD state attain the maximum power for a given $\mdot$, increasing their chance to reach the Bondi scale and produce feedback.
This suppresses the mass accretion rate to $\left< \mdot\right>/\mdotb= 1.5$\%, a staggering reduction by nearly 2 orders of magnitude from the peak, and the analytical \cite{bon52} expectation, \mdotb. 
The average jet power, $\left< L_{\rm j}\right>/ \mdotb c^2\simeq 2.9$\%, translates into extremely high jet efficiency, $\left< \eta\right> \simeq 190\%$: this implies that the \BH energy output in the form of jets exceeds the energy input in the form of the rest-mass energy. This is typical for rapidly spinning ($a\ge 0.9$) MAD \BH{}s, whose rotational energy is extracted -- by the continuous winding of magnetic fields on the event horizon -- faster than the accretion can replenish it \citep{tchekhovskoy2011efficient,2012MNRAS.423L..55T,mtb12}. 
Figure~\ref{fig:panel_fig}(b) shows that at $t\gtrsim65$k the normalized magnetic flux drops below the characteristic MAD value, $\phibh = 50$, the flow exits the MAD state, and the jets become progressively weaker (Fig.~\ref{fig:panel_fig}c) until their complete destruction at the event horizon at $t=78$k, when $\eta \le 0.1\%$ (Fig.~\ref{fig:panel_fig}d). With no strong jets to obstruct the accretion, the mass accretion rate (Fig.~\ref{fig:panel_fig}a) increases from $\mdot/\mdotb \sim1.5$\% in the MAD state to $\mdot/\mdotb\lesssim 10$\% in the SANE state.

\begin{figure*}[!ht]%
    \centering
    \hfill\includegraphics[width=0.48\linewidth]{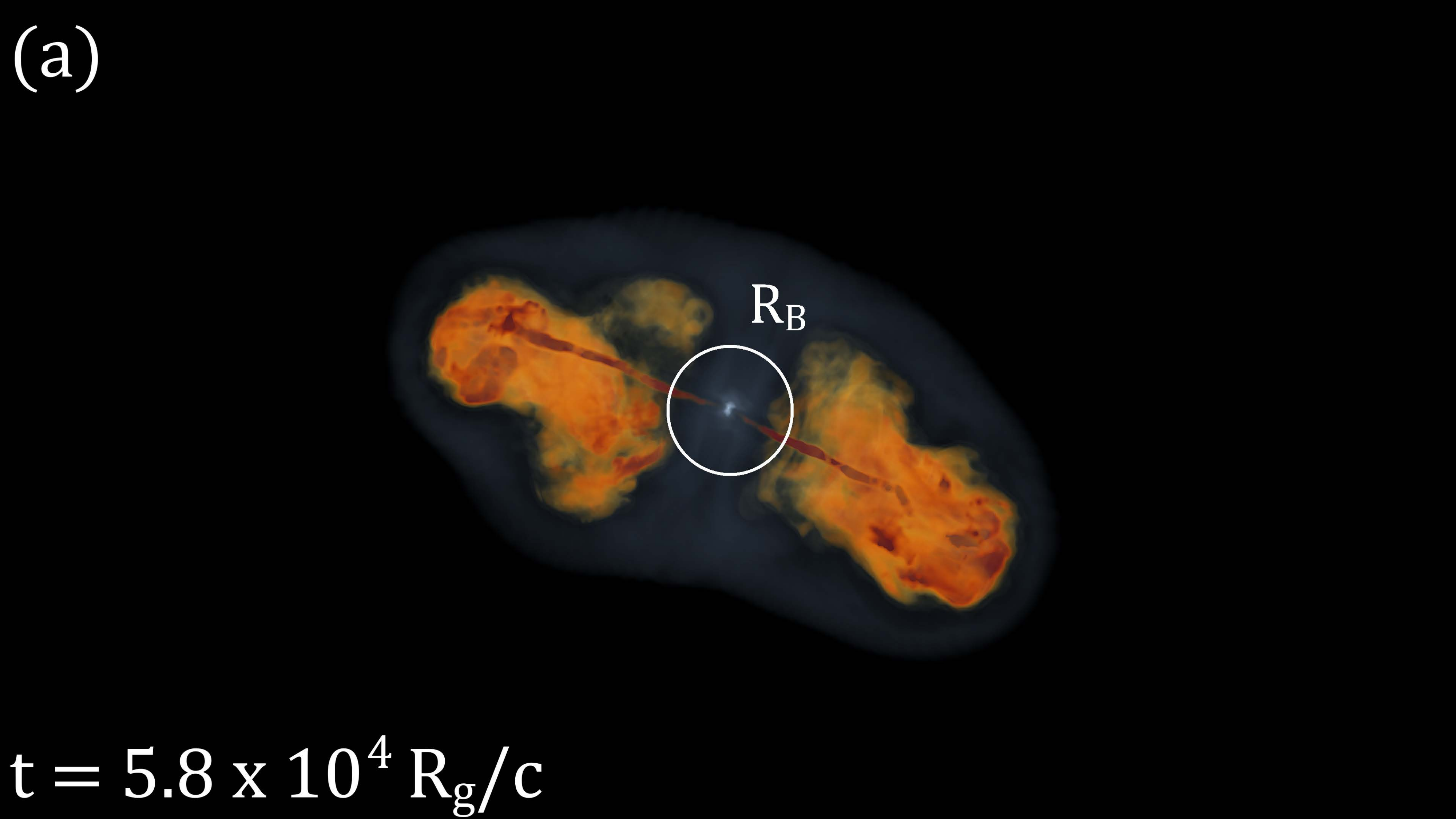}\,%
    \includegraphics[width=0.48\linewidth]{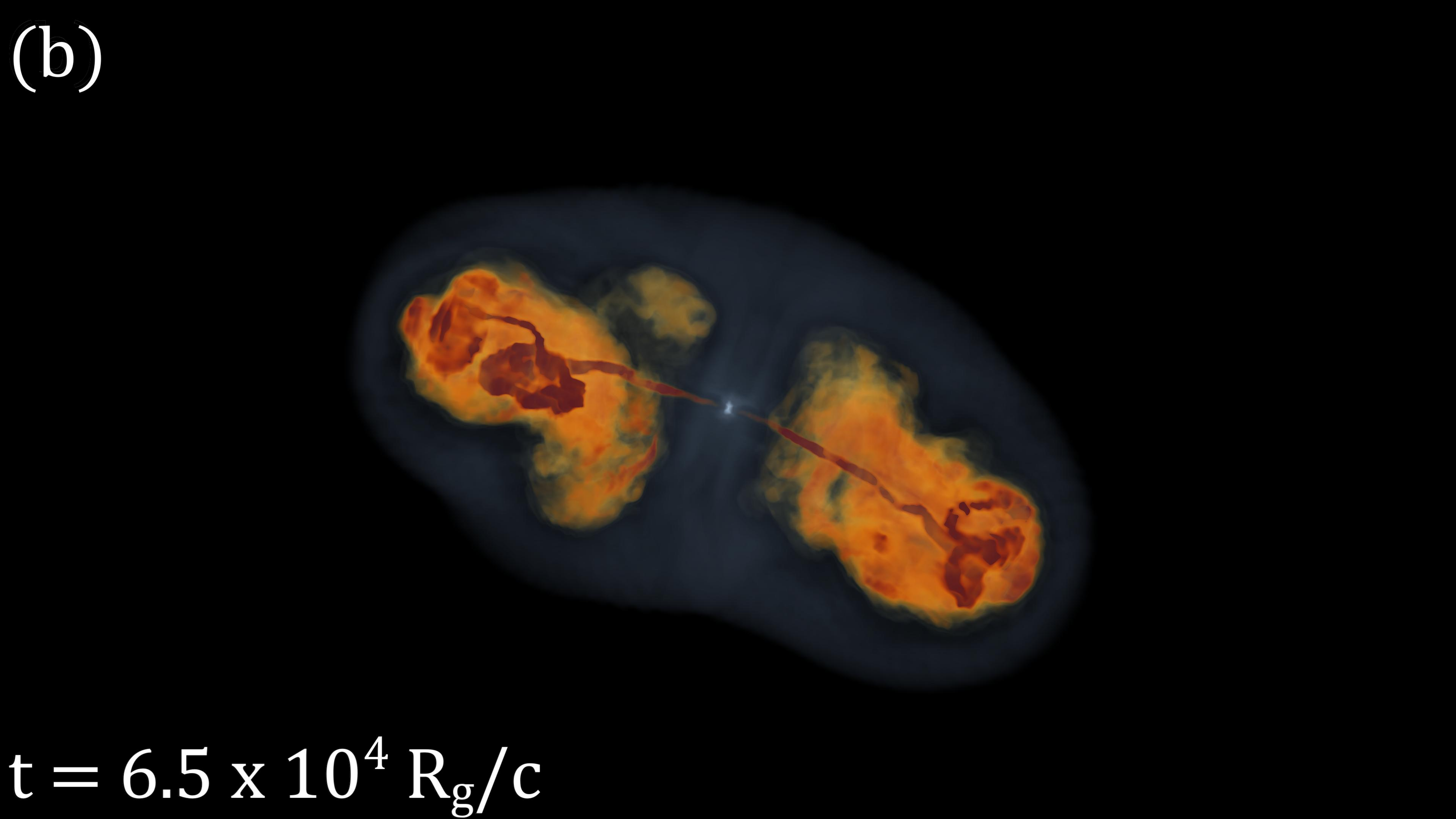}\hfill\hbox{}\\
    \hfill\includegraphics[width=0.48\linewidth]{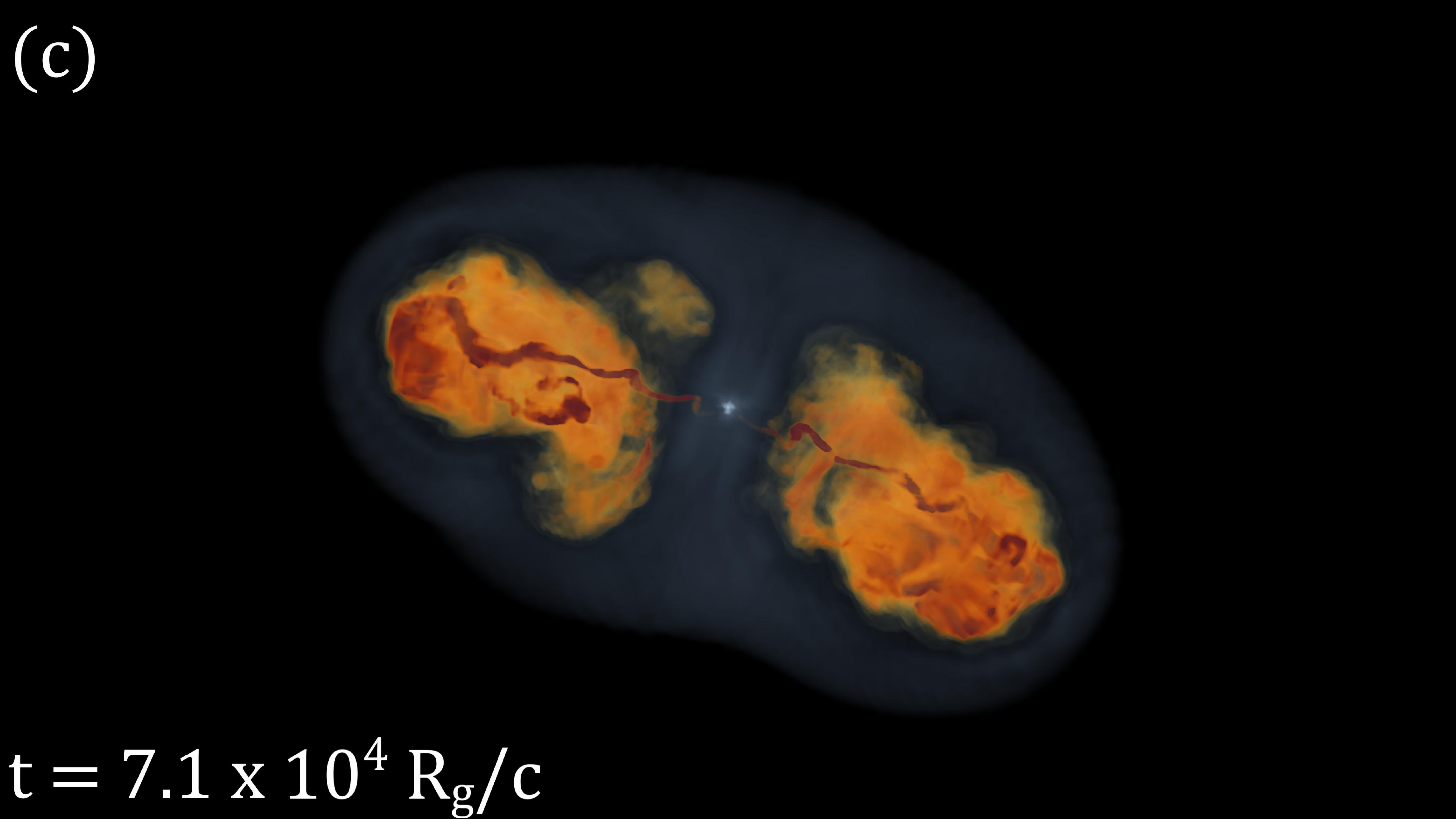}\,%
    \includegraphics[width=0.48\linewidth]{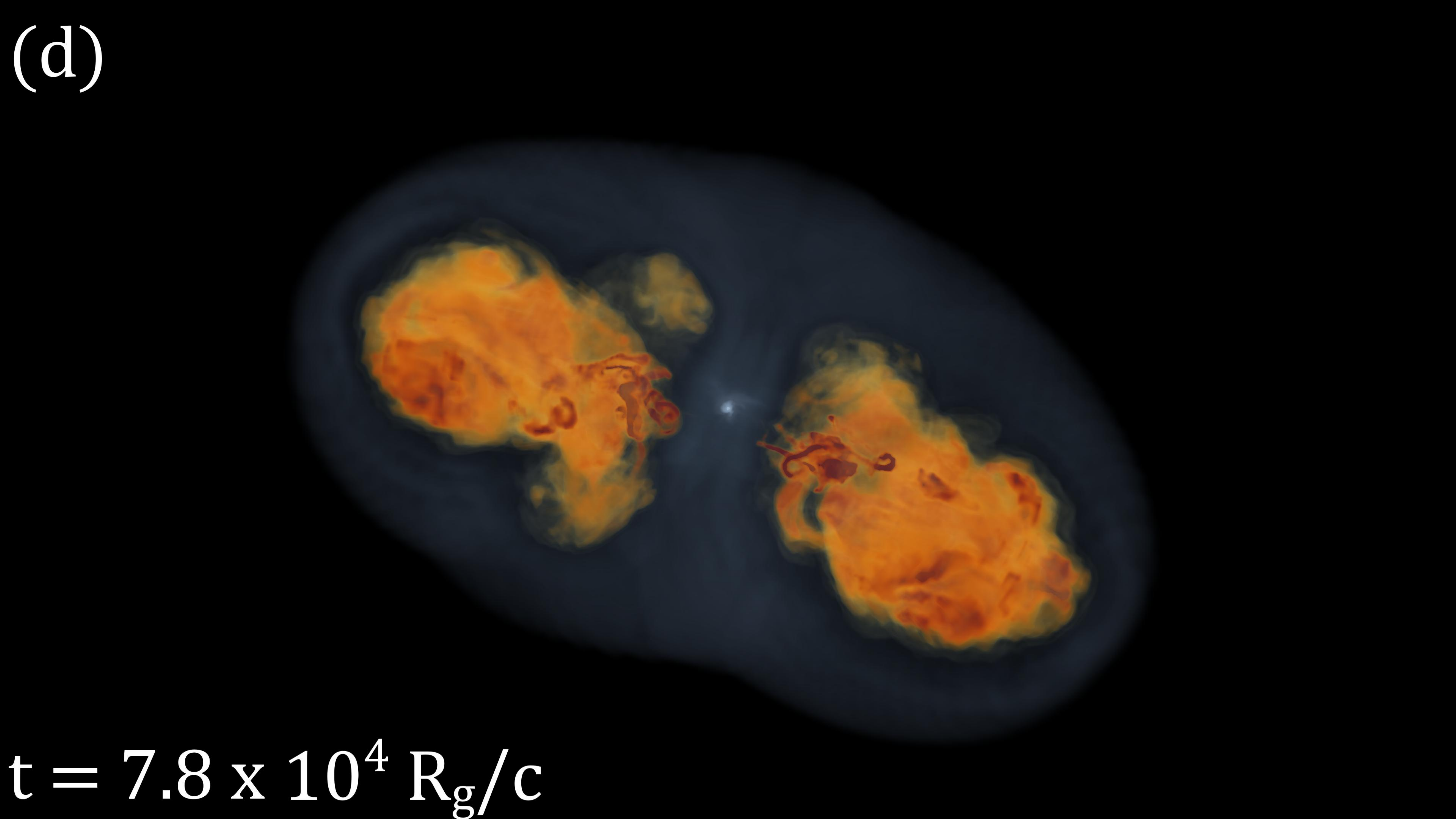}\hfill\hbox{}\\
    \hfill\includegraphics[width=0.48\linewidth]{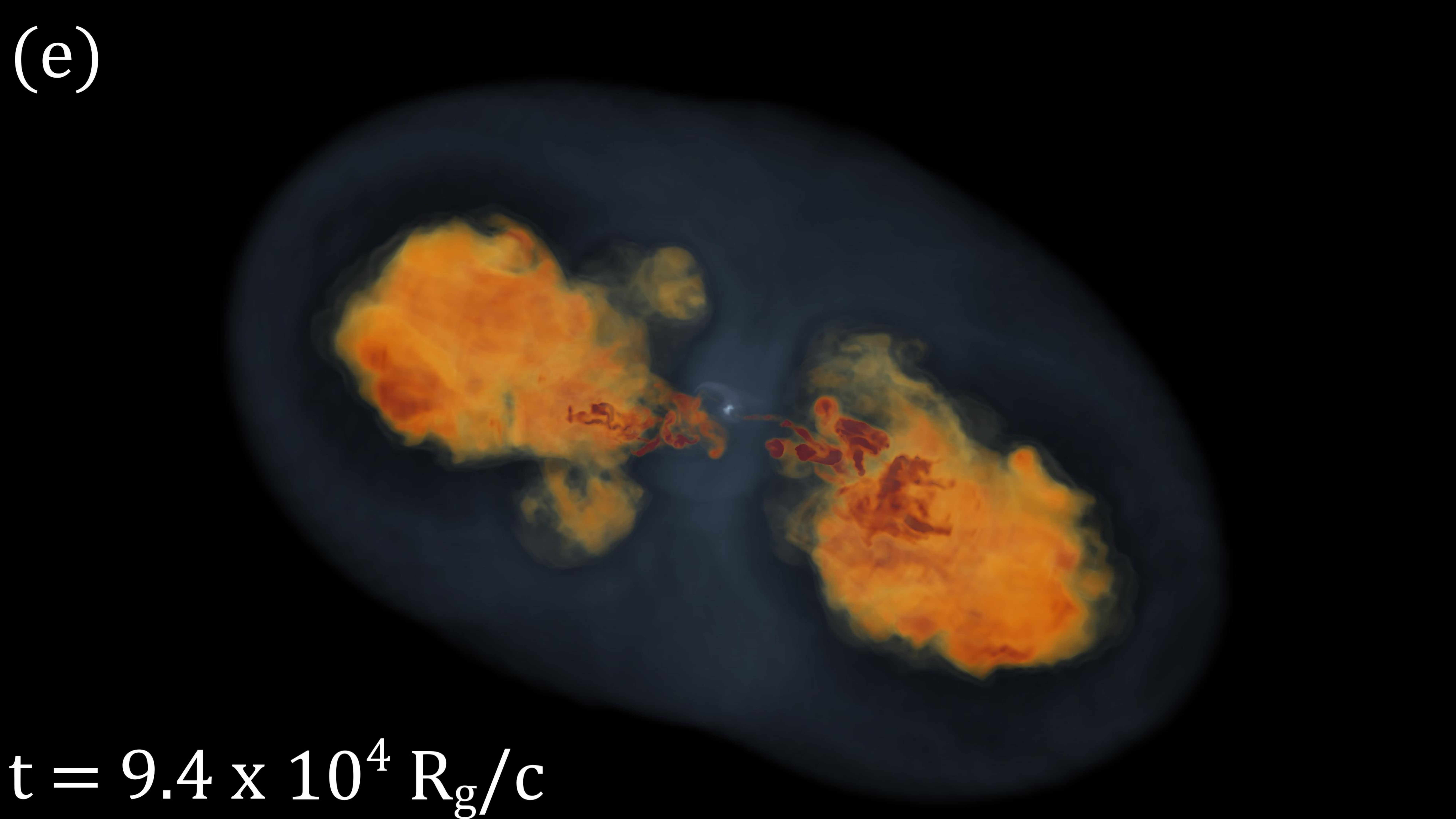}\,%
    \includegraphics[width=0.48\linewidth]{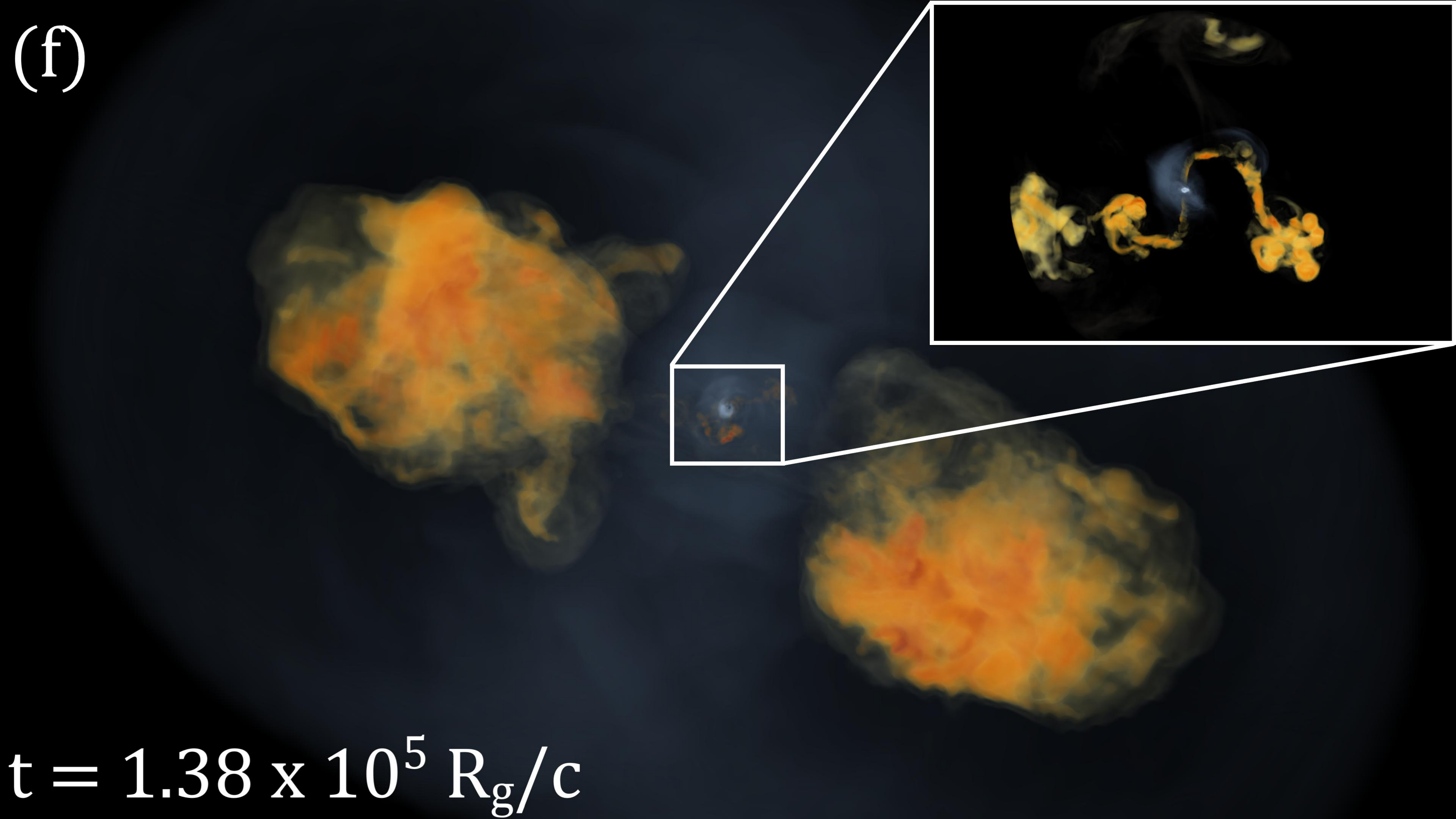}\hfill\hbox{}
    \caption{Three-dimensional volume renderings of density at different times show that as jets propagate out of the Bondi sphere, they become kink unstable and globally fall apart. The BH spin is pointing in the direction of jet propagation to the right in panel (a). A movie is available in Supplementary Information and on \href{https://www.youtube.com/watch?v=RumlBxHk3WQ&list=PLrqKzywvCPji5cufLwDzOgSMXcPLtE6T3&ab_channel=AretaiosLalakos}{YouTube (link)}. \textbf{[panel~a]:} The highly powered MAD jets (red) easily escape the Bondi sphere without obvious signs of instability, except near the head. At the contact points (hotspots), at which the jets drill through the ambient gas, the shocked jet material spills sideways, creates backflows, and forms the cocoons (yellow). The relativistic motion of the jets and cocoons shocks the ambient gas via bow shocks (blue). \textbf{[panel~b]:} As the jets propagate further out, this leaves enough time for the kink instability to grow, and it leads to helical bends near the hotspot. \textbf{[panel~c]:} The jet power decreases by $\sim 4$, and the kink unstable regions move in closer to the \BH. The decrease in power propagates through the jets faster than the cocoons can react, and the weakened jets get squeezed by the cocoons. \textbf{[panel~d]:} The central engine shuts off, and the jets break apart globally. \textbf{[panel~e]:}  Weakened jets wobble and lose focus: without consistent energy supply from the jets, mildly-magnetized cocoons become relics, which buoyantly rise outwards. Meanwhile, the wobbling jets transiently form and fall apart, without an opportunity to reach the Bondi sphere. \textbf{[panel~f]:}  Intermittent and weaker jets form, but get deflected sideways, as the accretion disk tilt continuously changes. The inset image shows that the jets develop extreme bends inside the Bondi radius.}
    \label{fig:kink_panels}%
\end{figure*}

\section{A Twisted Interlude} \label{sec:kink}
In this section, we explore the large-scale stability and propagation of the jets, and how they reach and impact the \BH feeding scales at $r\gtrsim \rb$.
Some jets can traverse vast distances, sometimes larger than the galaxy itself (e.g., Cygnus A, \citealt{1984ApJ...285L..35P}), whereas others appear to perish within the galaxy (e.g., M87,  \citealt{1991AJ....101.1632B}). This involves a scale separation of $\sim$10 orders of magnitude from the \BH to the outskirts of the galaxy. Along the way, the jets survive in the face of many obstacles, each of which can become a major dissipation-triggering event (e.g. internal, external, and/or recollimation shocks, MHD instabilities, magnetic reconnection, etc.), where a significant fraction of the jet energy can transform into radiation, and the jet can become locally or globally disrupted. Indeed, magnetized jets are prone to current-driven instabilities, with the kink instability being perhaps the most destructive \citep{bateman1978mhd,1992SvAL...18..356L,1993ApJ...419..111E,1998ApJ...493..291B,1999MNRAS.308.1006L,nlt09}.
Kink instability acts on the jet core, dislodging it from its original location and twisting it into a helical structure. In ambient media with a flat density profile, $\rho \propto r^{-\alpha}$, where $\alpha<2$, the kink instability is inevitable 
\citep[hereafter \citetalias{bromberg2016}]{bromberg2016}:
in such a medium, a jet with a constant opening angle would displace progressively more and more gas as it propagates out. The jet responds by becoming progressively more collimated. As we will see below (Sec.~\ref{sec:jet-stability-criterion}), this makes it more susceptible to the kink instability that can cause the jets to fall apart.

To visualize the large-scale morphology and evolution of the jets, we have created a 3D volume-rendered movie. We present a sequence of key snapshots in Fig.~\ref{fig:kink_panels}, taken at the times indicated by the vertical lines in Fig.~\ref{fig:panel_fig}. 
The jets are launched at $t=24$k (Fig.~\ref{fig:panel_fig}c), and by $t \simeq 28$k they reach the Bondi sphere and start interacting with the ambient medium (not shown in Fig.~\ref{fig:kink_panels}). Figure~\ref{fig:kink_panels}(a) shows that by $t = 58$k, the high-power MAD jets (red) drill their way through the ambient medium out to $r\simeq 4000\rg = 4\rb$. As the jets ram into the ambient medium, they form strong twin bow shocks at the jet heads. The shocks increase the internal energy at the heads, which typically renders these regions observable as hotspots. At the head, jet material spills sideways and flows back to create the \emph{inner cocoon} (orange), and the strong bow shock, caused by the jet's relativistic motion, heats up the ambient gas to create the \emph{outer cocoon} (blue). The jets are mostly straight apart from the bends in the outer half of the jets ($r \gtrsim 2000\rg$).
Figure~\ref{fig:kink_panels}(b) shows that at later times, $t=65$k, these bends become more pronounced, e.g., one of them twists the jet into a knot, as seen at $r\simeq2000\rg$ in the left jet.

Soon after this, the MAD state comes to an end: by $t=71$k, the jet power drops by a factor of $\sim 4$ (Fig.~\ref{fig:panel_fig}c). Figure~\ref{fig:kink_panels}(c) shows that the jets become more helical: the kink instability affects a larger fraction of the jet length, with the first jet bend showing up already at $r\lesssim1000\rg$.
Figure~\ref{fig:kink_panels}(d) reveals that by $t = 78$k the jets get globally destroyed: at this time,  the power of the jets essentially vanishes (Fig.~\ref{fig:panel_fig}c), and they no longer focus their energy into the twin cocoons. At later times, Figure~\ref{fig:kink_panels}(e)--(f) shows that short-lived jets still form close to the \BH but fall apart around, or even inside, the Bondi radius, and starve the cocoons of energy. With the energy supply via the jets mostly shut off, the mildly magnetized cocoons become relics, with a combination of leftover momentum and buoyancy, and occasional mergers with smaller cavities inflated by the wobbling jets, driving them outwards. At large scales, although weakened, bow shocks are still present and outrun the cocoons while injecting energy and momentum into the ambient gas. 

\subsection{Jet Stability Criterion}
\label{sec:jet-stability-criterion}
To quantitatively analyze the development of the kink mode in the jets, we use eqs.~\eqref{eqn:RM_fl} and \eqref{eqn:EM_fl}--\eqref{eqn:KE_fl} to express the ratio of total to rest-mass energy fluxes,
\begin{equation}
\mu  \equiv \frac{f_{\rm TOT}}{f_{\rm RM}} = -u_{t} (1 + h+\sigma),
\label{eq:mudef}
\end{equation}
which gives the maximum Lorentz factor a fluid element can attain if all of its EM and thermal energy fluxes were converted into the kinetic energy flux. To obtain eq.~\eqref{eq:mudef}, we used the facts that $f_{\rm TOT} = f_{\rm KE} + f_{\rm TH} + f_{\rm EM}$ and 
\begin{equation}
-u_t = \frac{f_{\rm KE}}{f_{\rm RM}} \approx \gamma \\, 
\label{eq:ut}
\end{equation}
and introduced the relativistic gas enthalpy per unit mass,
\begin{equation}
w \equiv \frac{f_{\rm KE}+f_{\rm TH}}{f_{\rm KE}} = 1+ \frac{u_{\rm g} + p_{\rm g}}{\rho c^2} = 1 + h \\, 
\label{eq:hdef}
\end{equation}
where $h$ is the non-relativistic gas enthalpy per unit mass (i.e. without the rest-mass contribution), and magnetization,
\begin{equation}
\sigma \equiv \frac{f_{\rm EM}}{f_{\rm KE}} \approx \frac{b^2}{4\pi\rho c^2} \\,
\label{eq:sigmadef}
\end{equation}
where the approximate equalities in eqs.~\eqref{eq:ut} and \eqref{eq:sigmadef} are accurate at large distances from the BH  \citep{2019MNRAS.490.2200C}. Below, unless stated otherwise, we will use the right-hand sides of eqs.~\eqref{eq:mudef}, \eqref{eq:hdef}, and \eqref{eq:sigmadef} as expressions for $\mu$, $h$, and $\sigma$, respectively.

Jets consist of an inner core of cylindrical radius, $R_{\rm c}$, dominated by the poloidal comoving magnetic field component, $b_{\rm p}$, and an outer region, dominated by the toroidal comoving field component, $b_{\rm tor}$~\citepalias{bromberg2016}. Here, $R_{\rm c} = r \rm sin\theta_c$ corresponds to the opening angle of the jet core, $\theta_{\rm c}$, at distance $r$, and we measure both $b_{\rm p}$ and $b_{\rm tor}$ in the fluid frame (see Appendix \ref{App:comoving_fields}). The kink instability growth timescale for a local fluid element is approximately the time it takes for an  Alfv\'en wave to travel around the circumference of the jet at the Alfv\'en velocity,
\begin{equation}
  \label{eq:va}
  v_{\rm A} = c \sqrt{\frac{\sigma}{1+h+\sigma}}.
\end{equation}
We define the kink instability growth time scale, as measured in the lab frame:
\begin{equation}
t_{\rm kink}  \equiv \eta_{\rm kink} \frac{2 \pi R \gamma}{v_{\rm A}} \times \cfrac{b_{\rm p}}{b_{\rm tor}}  \simeq \eta_{\rm kink}\cfrac{2 \pi R_{\rm c} \gamma}{v_{\rm A}}\,,
\label{eqn:growth_time0}
\end{equation}
where we used the jet Lorentz factor, $\gamma$, to convert the timescale from the fluid-frame to the lab-frame.
The ratio, $b_{\rm p}/b_{\rm tor}$, is the magnetic winding, and $R \times b_{\rm p}/b_{\rm tor}$ is the pitch, where $R=r \rm sin\theta$ is the cylindrical radius of the jet location in question. The factor, $\eta_{\rm kink} \simeq 5-10$, is the poorly constrained prefactor that enters the kink timescale for the jet to become considerably deformed \citep{Mizuno09, Mizuno12, Bromberg19}. The approximate equality in eq.~\eqref{eqn:growth_time0} evaluates the timescale at the edge of the jet core, where by definition, $b_{\rm p}=b_{\rm tor}$, and the value of the pitch is simply $R_{\rm c}$.
The kink instability develops in the fluid frame, and the available time for its growth is the propagation time of the jet fluid from the \BH, moving at velocity $v$. In the lab frame, this dynamical time is: 
\begin{equation}
t_{\rm dyn} = \int_0^r \cfrac{\der{r'}}{v} \simeq \cfrac{r}{v}\,, 
\label{eqn:dyn_time}
\end{equation}
where we assume constant jet velocity, $v$, ignoring the initial acceleration of the jet, and that the jet travels only in the radial direction (not true in a bent jet).
The ratio of the kink to dynamical timescale (eqs.~\ref{eqn:growth_time0} and~\ref{eqn:dyn_time}) gives us the stability parameter:
\begin{equation}
\Lambda = \eta_{\rm kink} \cfrac{2 \pi R_{\rm c} \gamma v }{v_{\rm A} r}\,.
\label{eqn:lambda}
\end{equation}
    The jet is stable at $\Lambda > 2$, and unstable at  $\Lambda < 2$ \citepalias{bromberg2016}. Calculating the radius of the jet core is a non-trivial task since the jet core can displace itself from the $z$-axis in a non-axisymmetric fashion. To compute $R_{\rm c}$, we first evaluate the solid angle subtended by the jet core, $\Omega_{\rm c}$, which we identify as satisfying both $\mu>3$ and $b_{\rm p} \ge b_{\rm tor}$ conditions. Then, we can calculate the opening angle from $\Omega_{\rm c}=2\pi (1-{\rm cos}\theta_{\rm c})$, and then $R_{\rm c}=r {\rm sin}\theta_{\rm c}$.
    Without loss of generality, we focus on one of the twin jets, the one pointing in the direction of the \BH spin, $\vec a$.

\begin{figure*}[t] 
\centering
\includegraphics[width=1\textwidth]{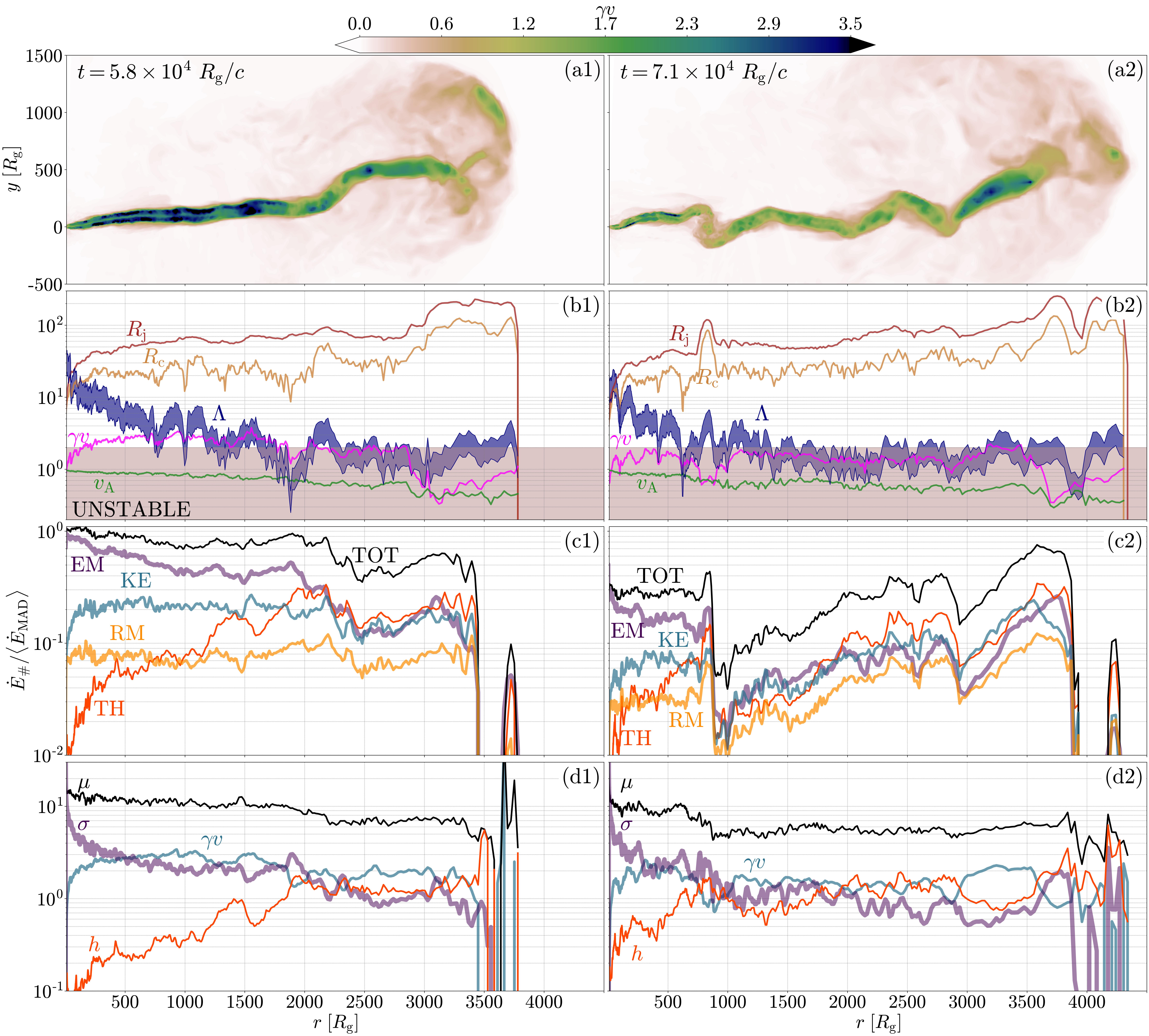}
\vspace{-5mm}
\caption{
Kink instability disrupts the jets as their power drops four-fold upon the exit out of the MAD state at $t\approx65$k.
{\bf [panels~(a)]:} Transverse projection of $\sigma$-weighted proper velocity, $\gamma v$, reveals large-scale jet bends, telltale signs of the kink instability (brown shows low and blue high values, see color bar).  
%
{\bf [panels~(b)]:} Whereas at the earlier time ($t=58$k, panels~a1 and b1) only the outer jet, $r\gtrsim2000\rg$, shows bends and satisfies the kink instability criterion, $\Lambda\lesssim2$, at the later time ($t=71$k, panel~a2 and b2) most of the jet, $r\gtrsim800\rg$, shows bends and satisfies the criterion. Jet bends cause the core-average proper jet velocity $\gamma v$ (magenta) to significantly decrease from $\gamma v\simeq 3$ (panel~b1) to $\gamma v\lesssim 2$ (panel~b2) and result in a comparable decrease in $\Lambda$ (eq.~\ref{eqn:lambda}). Although the jet radius, $R_{\rm j}$ (dark red), also decreases, this noticeably affects neither the core radius, $R_{\rm c}$ (gold), nor core-average Alfv\'en velocity, $v_{\rm A}$ (green). 
%
{\bf [panels~(c)]:} Angle-integrated fluxes, or power, through the jet ($\mu>3$) normalized to the time-average total MAD power, $\langle \dot{E}_{\rm MAD}\rangle$. The normalized total jet power, $\dot{E}_{\rm TOT}$ (black),  remains approximately constant in stable jet regions. The kinetic energy component, $\dot{E}_{\rm KE}$ (light blue), gradually increases at the expense of the EM component, $\dot{E}_{\rm EM}$ (purple), which dominates the total energy power until the instability sets in at $r\gtrsim 2000\rg$ in panel (c1) and $r\gtrsim 800\rg$ in panel (c2). Panels (a1) and (a2) show that in the unstable regions, the jet develops one or more bends. Whereas just before each substantial bend the thermal component (orange-red) grows up to equipartition with the EM component (purple), $\dot{E}_{\rm TH}\sim \dot{E}_{\rm EM}$, after each bend all energy power components drop, followed by a gradual increase until the next bend (e.g., for bends at $r/\rg \simeq 800, 1300, 2000, 2500$, and $3000$ in panel c2): we attribute this to the ``accordion'' effect -- the jet bends allow the jets to elongate and rarefy longitudinally. The rest-mass power  $\dot{E}_{\rm RM}$ (orange) remains subdominant to the rest of the components.
{\bf [panels~(d)]:} Jet-average proper velocity, $\gamma v$ (light blue, obtained via $\gamma\equiv \dot{E}_{\rm KE}/\dot{E}_{\rm RM}$), increases as the jets accelerate at the expense of the decreasing jet-average magnetization, $\sigma\equiv\dot{E}_{\rm EM}/\dot{E}_{\rm KE}$ (purple), before decelerating at jet bends. Jet-average enthalpy, $h\equiv \dot{E}_{\rm TH}/\dot{E}_{\rm KE}$ (orange-red), increases at jet bends due to energy dissipation. The $\mu\equiv \dot{E}_{\rm TOT}/\dot{E}_{\rm RM}$ parameter (black) decreases due to the jet losing its energy to the ambient medium (via ambient gas displacements and shocks). 
 }
\label{fig:quant_pan}
\end{figure*}

\subsection{Jet Bends Set the Speed Limit}
\label{sec:jet-speed-limit}
Figure~\ref{fig:quant_pan}(a1) shows a transverse projection of the jet proper velocity, $\gamma v$, at $t=58$k (the same time as in Fig.~\ref{fig:kink_panels}a). Here, we average $\gamma v$ along the line of sight (which is perpendicular to the direction of \BH spin, $\vec{a}$), and weigh it with the magnetization, $\sigma$, to highlight the internal structure of the jet. At this time, most of the jet is free of significant bends, except towards the head, at $r\gtrsim2000\rg=2\rb$. As the jet bends, its proper velocity drops from $\gamma v\simeq 3$ (dark blue) to $\gamma v \simeq 1.7$ (green), which corresponds to $\gamma \simeq 2$: the slower the jet, the easier for it to make sharp turns (see below). 
The jet maintains its transrelativistic proper velocity until it reaches the head  (hotspot), splashes sideways and backward, and creates the backflows (brown)
at $\gamma v \lesssim 1$. 
Figure~\ref{fig:quant_pan}(a2) shows the jet after the MAD state has ended, at $t=71$k (same time as in Fig.~\ref{fig:kink_panels}c). By this time, the average jet power has decreased by a factor of $\sim 4$, compared to Fig.~\ref{fig:quant_pan}(a1). Weaker jets are less rigid, thus their interaction with the ambient medium can bend them more easily. Indeed, the jet develops bends that are much stronger and closer to the \BH than in the MAD state. These bends force most of the jet to become mildly relativistic, $\gamma v \le 1.7$. Even tiny, visually imperceptible, bends can have dramatic consequences for relativistic jets. 

For relativistically-magnetized jets, the consequences can become particularly dire when the jets become super-fast magnetosonic, i.e., when they outrun their own fast magnetosonic waves, which for cold flows translates to a proper velocity, $u_{\rm F} = \gamma_{\rm F}v_{\rm F} \approx \sigma^{1/2}$ \citep{tch10b}. If we introduce the fast Mach number, $M_{\rm F} = \gamma v/u_{\rm F} \approx \gamma v/\sigma^{1/2}$, we can write the fast wave Mach cone opening angle as
\begin{align}
    \theta_{\rm Mach}
    \equiv  \sin^{-1}\frac{1}{M_{\rm F}} \approx \frac{u_{\rm F}}{\gamma v} \approx \frac{\sigma^{1/2}}{\gamma v},
\end{align}
where the first approximate equality applies when the jets are in the highly super-fast magnetosonic regime, $M_{\rm F}\gg 1$, and the latter applies in the cold limit.
Why do relativistic jets have a hard time making turns? This is because in order for a jet to navigate a turn smoothly, the jet must be able to anticipate that the turn is coming. Here, $\theta_{\rm Mach}$ gives the jet's ``field of vision'' angle, i.e., the maximum angular deviation within which fast waves can mediate the total pressure perturbations. Thus, the jets cannot bend by more than the fast Mach cone opening angle, 
\begin{align}
\theta_{\rm bend} 
    &\lesssim \theta_{\rm Mach}
    \approx  \frac{u_{\rm F}}{\gamma v}
    \approx  \frac{\sigma^{1/2}}{\gamma v},
    \label{eq:theta_bend}
\intertext{
or, equivalently, the jets cannot exceed the ``speed limit'', $u_{\rm max}$,
}
\gamma v & \lesssim u_{\rm max} \equiv \frac{u_{\rm F}}{\theta_{\rm bend}}  \approx \frac{\sigma^{1/2}}{\theta_{\rm bend}},
\end{align}
where the latter approximate equality applies in the cold limit.
The stronger the bend (larger $\theta_{\rm bend}$), the smaller the speed limit, $u_{\rm max}$, which is the maximum allowed jet proper velocity, $\gamma v$, for which the jets avoid the development of internal shocks. 
If $\gamma v$ exceeds $u_{\rm max}$, the jets develop oblique internal shocks that discontinuously reduce the jet velocity. 
Bends are not the only mechanism through which jets can develop shocks and decelerate. For instance, if the jets conically expand into a medium, they accelerate rapidly and soon become super-fast magnetosonic. At some point, the pressure of the hot cocoons engulfing the jets starts to dominate over the jet internal pressure and squeezes the jets. As a result, the jets develop collimation shocks behind which the jet material slows down. Magnetic stresses operating downstream of the shocks introduce additional compression forces that cause the jets to pinch and form narrow nozzles in which the kink instability can grow more efficiently and dissipate energy \citep[e.g.][]{lyub09, bromberg2016, barniol_duran2017}.

\subsection{What Triggers Kink Instability in Our Jets?}
\label{sec:what-triggers-kink}

Figure~\ref{fig:quant_pan}(b) shows the radial profiles of proper, $\gamma v$ (magenta), and Alv\'en, $v_{\rm A}$ (green), velocities, which we have angle-averaged over the jet core of radius $R_{\rm c}$ (shown in gold, see Sec.~\ref{sec:jet-stability-criterion} for definition). For comparison, we also show the radius of the jet, $R_{\rm j}$ (dark red). 
Using eq.~\eqref{eqn:lambda}, we can now compute the stability parameter, $\Lambda$ (blue): here, the line thickness indicates the uncertainty range of $\eta_{\rm kink}=5-10$ in the definition of the stability parameter. That $\Lambda \gtrsim 2$ at $r\lesssim 1500\rg$ in Figure~\ref{fig:quant_pan}(b1) implies that according to the stability criterion the jet is expected to be stable: Fig.~\ref{fig:quant_pan}(a1) shows that at these distances, the jet appears to be mostly straight. At larger radii, however, we have $\Lambda\lesssim2$, suggesting an instability: consistent with this expectation, the jet develops bends at $r \gtrsim 1500\rg$. At the later time, Fig.~\ref{fig:quant_pan}(b2) shows that $\Lambda\lesssim2$ throughout most of the jet, already at $r\gtrsim500\rg$, indicating that according to the stability criterion, we expect that most the jet is susceptible to the kink instability. Fig.~\ref{fig:quant_pan}(a2) shows multiple strong bends throughout the length of the jets. We note that strictly speaking, eq.~(\ref{eqn:lambda}) is only valid for small perturbations (i.e., in the limit of small jet bends) and not applicable once the bends become extreme (see Sec.~\ref{sec:ext-vs-int-kink}).

Figure~\ref{fig:quant_pan}(b1) shows that in the stable region, at $r\lesssim 1500\rg$, the jet core is relativistic, $\gamma v \simeq 3$, with an average radius of $R_{\rm c} \simeq 20 \rg$. The multiple dips in $\gamma v$ correspond to recollimation or oblique shocks in the jet, at $r/\rg\simeq 800, 1300,$ and $1500$. At these locations, we see that the velocity drops to $\gamma v \simeq 2$, and 
the jet total and core radii dip; Figure~\ref{fig:quant_pan}(a1) shows that the jet tends to develop bends at the same locations. 
At $r \gtrsim 2000\rg$, the jet displays its largest bend: at the beginning of the bend, the jet velocity dips to $\gamma v \simeq 1.5$ and the core radius dips to $R_{\rm c}\lesssim10\rg$. As the jet bend proceeds, the core radius gradually increases to $R_{\rm c}=50\rg$ likely reflecting the dissipation of the toroidal field; the velocity also recovers and levels off at $\gamma v \gtrsim 2$. Neither the jet radius $R_{\rm j}$, which remains $\sim 3-4$ times wider than the core, nor the Alfv\'en speed, which remains roughly constant at $v_{\rm A}/c \sim 0.5-0.7$, appear to be affected by the bend. 

At the later time, Fig.~\ref{fig:quant_pan}(b2) shows that the jet core moves at a transrelativistic speed, $\gamma v \lesssim 2$, and both $R_{\rm c}$ and $\Lambda$ show multiple dips, likely associated with oblique shocks (e.g., at $r/\rg\simeq 400$ and $600$), at which the stability parameter dips down to an unstable level, $\Lambda<2$, indicating the presence of the instability. Figure~\ref{fig:quant_pan}(a2) shows that the jet disrupts at $r=800\rg$, and the velocity decreases to $\gamma v \lesssim 1$. At the same time, both $R_{\rm c}$ and $R_{\rm j}$ spike due to the $r = {\rm constant}$ cross-section slicing the jets at an angle and making jet features appear wider. Even at such high resolutions as ours, the simulation resolves the transverse extent of the jet core ($\sim 30\rg$) out to $r\simeq 1500\rg$ by 5 cells. The simulation resolves the transverse extent of the jet ($\sim 100\rg$) out to $r\simeq 5000\rg$ by 5 cells, although at such large distances, it might lose some details of its internal structure.

\subsection{Energy Partition and Dissipation}
\label{sec:dissipation}
Figure~\ref{fig:quant_pan}(c)--(d)
 shows angle-average quantities, over the \emph{entire} jet cross-section, which we define as the relativistic region, $\mu>3$, to avoid the contamination caused by the mildly magnetized cocoon and weakly magnetized ambient gas. (This is in contrast to Fig.~\ref{fig:quant_pan}b that considered averages over the jet core.)

Figure~\ref{fig:quant_pan}(c1) shows various components of the radial angle-integrated jet energy flux, essentially the energy rate, or power. We normalize all of the power components by the total power flowing through the entire jet cross-section ($\mu>3$) at $r=8\rg$, averaged over the duration of the MAD state, $\dot{E}_{\rm MAD} \equiv \langle \dot{E}_{\rm TOT}(r=8\rg)\rangle_{\rm MAD}$; here $\langle \dots \rangle_{\rm MAD}$ denotes the time-average over the MAD state, which we highlighted in purple in Fig.~\ref{fig:panel_fig}
\footnote{Because most of the EM energy flux out of the \BH flows through the highly magnetized jets, we have $\dot{E}_{\rm MAD} \approx 0.5 \langle L_{\rm j}\rangle_{\rm MAD}$, where the factor of $0.5$ is needed because $L_{\rm j}$ is the EM power of both jets.}.
The total jet power, $\dot{E}_{\rm TOT}$ (black line in Fig.~\ref{fig:quant_pan}c1), remains approximately constant at $r\lesssim 2000\rg$, and drops slightly thereafter, just as the jet starts to bend, as seen in Fig.~\ref{fig:quant_pan}(a1). 
The jet electromagnetic power, $\dot{E}_{\rm EM}$ (purple), dominates the jet power at $r \lesssim 2000\rg$, whereas the thermal component of the power, 
$\dot{E}_{\rm TH}$ (orange-red),
steadily increases until, at $r\sim2000\rg$, it reaches equipartition with 
$\dot{E}_{\rm EM}$, a tell-tale sign of energy dissipation \citepalias{bromberg2016}. The increase in 
$\dot{E}_{\rm TH}$
just as the jet starts to bend most likely comes from the dissipation of EM and kinetic energy into heat due to kink, mixing instabilities, and/or shocks caused by the bend (see Sec.~\ref{sec:jet-speed-limit}).\footnote{At much smaller distances, close to the jet base, $r\lesssim 100 \rg$, highly magnetized inner jet regions can heat up due to the numerical truncation error that affects the smallest energy scale: in this case, the internal energy; however, these effects tend to operate when the internal energy is very small, $\dot{E}_{\rm TH} \lesssim {\rm few}\times 0.01\dot{E}_{\rm TOT}$.} The kinetic energy power component, 
$\dot{E}_{\rm KE}$ 
(light blue), steeply rises at $r\lesssim 100\rg$, while the rest-mass power component  
$\dot{E}_{\rm RM}$
(orange) is subdominant compared to the other components.

At $r \lesssim 2000\rg$, 
$\dot{E}_{\rm KE}$ 
remains approximately constant because the jet becomes cylindrical (due to external collimation by the constant density ambient medium). However, once the jet becomes unstable to the kink instability and develops strong bends, $\dot{E}_{\rm KE}$ drops. At the later time, $t=71$k, Fig.~\ref{fig:quant_pan}(c2) shows that $\dot{E}_{\rm TOT}$ steeply drops at $r\simeq800\rg$, right where the jet makes a sharp $90-$degree turn, as seen in projection on Fig.~\ref{fig:quant_pan}(a2). In fact, all of the energy power components experience similarly sharp drops, since we are only accounting for the radial component of the fluxes, which vanishes. At the location of such drops, which correspond to strong bends, the jet is disrupted and the energy rate does not increase, meaning it is lost. This is anticipated since (i) when the jets fall apart, they can mix with the ambient medium, lower the specific energy, and thus, they won't be picked out by our jet-criterion $\mu\ge 3$, and (ii) the jet head is spending energy into displacing the surrounding gas, while also losing a fraction of its energy into the formation of the cocoons. 


In Fig.~\ref{fig:quant_pan}(d1) we plot the jet area-averaged radial profiles of $\mu$ parameter (black), the magnetization, $\sigma$ (purple), the proper velocity, $\gamma v$ (light blue), and specific gas specific enthalpy, $h$ (orange-red): To compute the jet-average values we expressed them through the ratios of jet power components (i.e. using $\dot{E}_{\#}$ in place of $f_{\#}$ in the definitions \ref{eq:mudef} -- \ref{eq:sigmadef}). Figure~\ref{fig:quant_pan}(d1) shows that the jet accelerates to relativistic velocities, $\gamma v \simeq 2$ at $r\lesssim 100\rg$ and $\gamma\simeq3$ at $r\lesssim1000\rg$. 
At $r\lesssim 2000\rg$ the jet is still magnetized, $\sigma \gtrsim 2$, the proper velocity remains (on average) relativistic, $\gamma v\gtrsim 3$, and the enthalpy, h, slowly increases, similar to $\dot{E}_{\rm TH}$. Once the jet becomes unstable and develops bends, the magnetic energy dissipates into heat, until
$\sigma \lesssim h$, i.e. until the magnetic energy comes into equipartition with the thermal energy. Kinetic energy is also reduced, with the jet decelerating down to the transrelativistic value $\gamma v \simeq 1.7$, as the jet needs to slow down in order to be able to navigate jet bends (see Sec.~\ref{sec:jet-speed-limit}). 
In Fig.~\ref{fig:quant_pan}(d2), we see a similar behavior as in Fig.~\ref{fig:quant_pan}(d1), but with the kink instability happening much closer to the \BH, already at $r=800\rg$. Both the magnetization and proper velocity drop, $\sigma \lesssim h$ and  $\gamma v\simeq 1.7$, and enthalpy $h$ increases. 

\begin{figure*}[!ht]%
\centering
\includegraphics[width=1\textwidth]{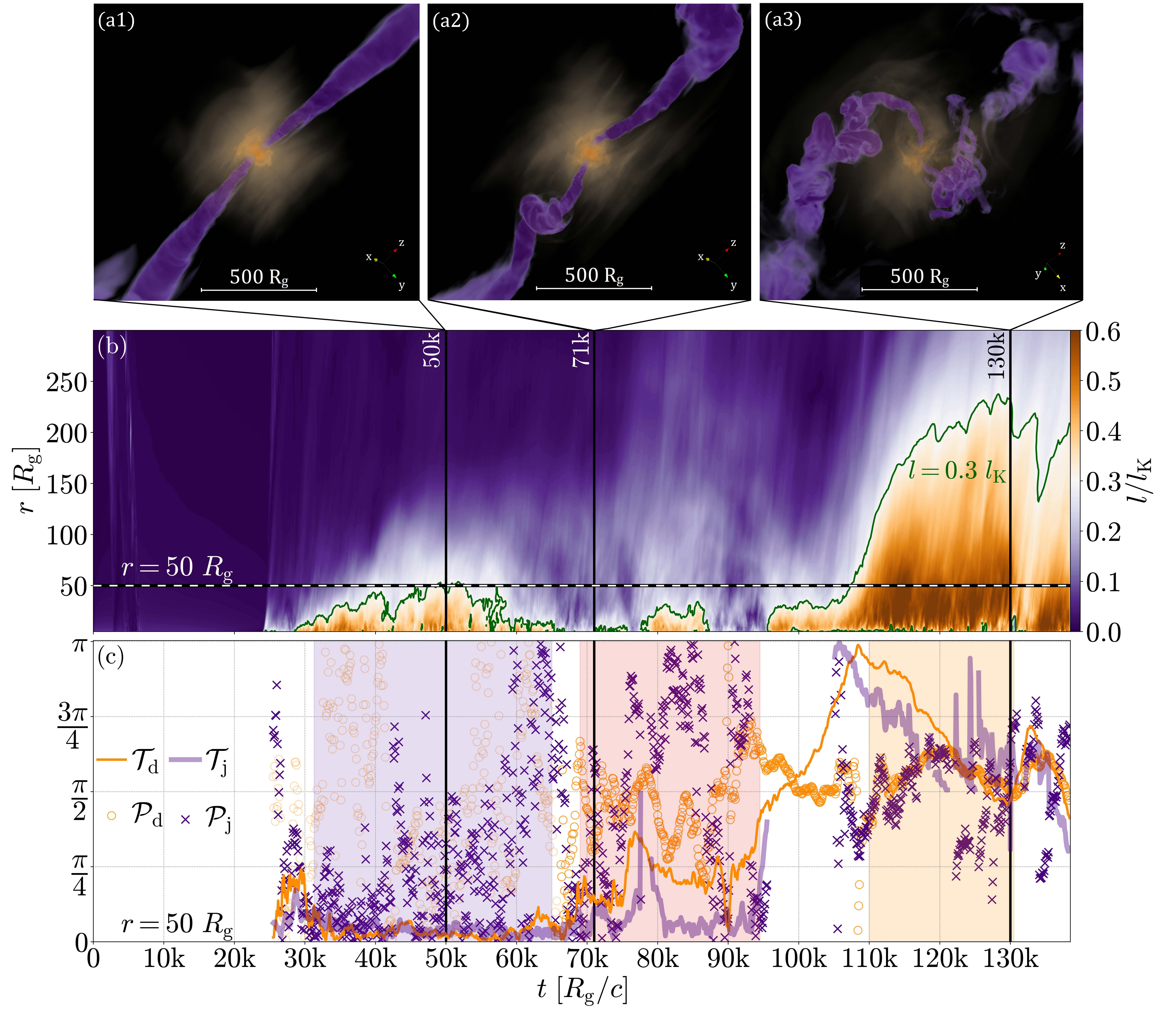}
\vspace{- 5mm}
\caption{
The morphology, strength, and directions of the angular momentum of the accretion flow during the MAD, \statetwo, and \statethree states. {\bf [panels~(a)]:} We show 3 distinct 3D volume-rendered images of density. Panel(a1) captures the system during the MAD state, at $t=50 \rm K$, where the jets (purple) follow the direction of the \BH's spin, and the accretion disk (orange) lies perpendicularly to the jets. Panel(a2) captures the system in the \statetwo state, at $t = 71$k, with weakened jets showing strong bends near the BH and getting significantly kinked. Panel(a3) captures the system in the \statethree state, at $t =135$k. The jets are on average 1 order of magnitude weaker and misaligned from the spin direction, The disk is also significantly tilted, at about $\pi/2$, and processing, causing significant bends and wobbles in the jets. {\bf [panel~(b)]:} We plot the density-weighted specific angular momentum normalized to the local Keplerian value in a space-time diagram. Angular momentum builds around the \BH promptly after freefall ($t_{\rm ff}\simeq 22$k) with an average disk of size $10-50 \rg$, chosen for $l \gtrsim 0.3 l_{\rm K}$. Between, $65\K \le t \le 100$k the accreted angular momentum forms intermittent and tilted disks with  $l \le 0.3 l_{\rm K}$ which results in intermittent and wobbly jets. After $t\simeq 100$k angular momentum starts building up, this time reaching larger radii $\simeq 200 \rg$, however, near the \BH the intermittent jets push most of the gas away while oscillating wildly, stopping the angular momentum from coherently adding up and stabilizing the jets. 
{\bf [panel~(c)]:} We show the tilt and precession angles of the disk and north jet, measured at a distance of $r=50\rg$. 
During the MAD state (purple shaded), both the tilt of the disk $\mathcal{P_{\rm d}}\lesssim \pi/16$ (orange) and jet $\mathcal{P_{\rm j}}\lesssim \pi/16$ (purple), as they are aligned with the BH spin. The disk precession angle (orange circles), $\mathcal{T_{\rm d}}$, is ill-defined as the disk tilts about the $z$-axis, while the jet precession angle (purple crosses), $\mathcal{T_{\rm d}}\simeq \pi/2$, which means it propagates at a preferential direction. Once we transition from the  MAD to the \statetwo state (red shaded), at $t\ge 65$k, the disk angular momentum has decreased, its tilt increases, reaching up to $\pi/2$, and rapidly precesses. The jet tries to follow and forms large bends (see Fig.~\ref{fig:angular_momentum}a2) that lead to its destruction. Between $95 \K \lesssim t \lesssim 110$k, the disk's tilt goes from $\pi/2$ to $\pi$, flipping to retrograde, while the jet is disrupted. During the \statethree state, at $t\ge110$k, the disk flips again, which coincides with the flipping of the magnetic flux on the Northern hemisphere, in Fig.~\ref{fig:panel_fig}(b). After $t\gtrsim 105\K$ we show the tilt and precession of the counter jet, as it attempts to follow the rocking motion of the disk.}%
\label{fig:angular_momentum}%
\end{figure*}

\section{A Turbulent Epilogue} \label{sec:epilogue}
Figures~\ref{fig:angular_momentum}(a) show 3D volume-rendered images of the low-density (purple) jet and the high-density accretion flow (orange). Figure~\ref{fig:angular_momentum}(a1) shows the system at $t=50$k, when it is in the MAD state: the jets are powerful and mostly bend-free, and the accretion flow near the BH shows a disk-like structure. Figure~\ref{fig:angular_momentum}(a2) shows the system after it exited the MAD state, where the jet power has decreased by a factor $\sim 4$. The weakened jets appear to have bends originating from wobbles near their launching region.
Figure~\ref{fig:angular_momentum}(a3), shows jets with extreme bends, misaligned from the BH spin ($z-$)axis. The accretion flow is also significantly misaligned, compared to Figs.~\ref{fig:angular_momentum}(a1, a2). This shows that the evolution, stability, and destruction of the jets are correlated with the underlying state of the accretion flow near the launching region.

To understand the evolution of the system, we calculate the density-weighted specific angular momentum, $\vec{l}=\vec{r} \times \vec{v}=(l_x, l_y, l_z)$. In particular, we compute the magnitude of $\vec{l}$:
\begin{equation}
l =  \frac{\iint \rho dA (l^2_x +l^2_y+l^2_z)^{1/2}}{ \iint \rho dA} ~.
\end{equation}
The angular momentum computed this way is appropriate in flat spacetime, $r\gtrsim10 \rg$. We normalize the resulting specific angular momentum with the specific Keplerian angular momentum $l_{\rm K}$ around a \BH of spin $a$ \citep{shapiro_black_holes_1986}:
\begin{equation}
    l_{\rm K}(r) = \cfrac{\sqrt{r} (r^2 -2a\sqrt{r}+a^2)}{r\sqrt{r^2-3r+2a\sqrt{r}}}~.
\end{equation}

Figure~\ref{fig:angular_momentum}(b) shows the magnitude of $l/l_{\rm K}$ on a spacetime diagram. The three vertical black lines indicate the times of the 3D panels in Fig.~\ref{fig:angular_momentum}(a).
We choose $l/l_{\rm K} \simeq 0.3$ as a fiducial value to separate the accretion disk from the rest of the gas, which we overplot as a green contour in Fig.~\ref{fig:angular_momentum}(b). Although initially the gas had zero angular momentum, within several \rg from the \BH the frame-dragging \citep{1918PhyZ...19..156L} due to the \BH high spin ($a=0.94$), combined with large-scale magnetic fields, to force the accretion flow to partially corotate with the \BH, and the flow develops an azimuthal angular momentum component \citepalias[see also][]{ressler2021magnetically}. Additionally, the MRI excites disk turbulence and transports the angular momentum outwards, so the disk can grow in size. Indeed, by $t \simeq 50$k, the accretion disk size has grown up to $50 \rg$. 


Figure~\ref{fig:angular_momentum}(a1) shows that once formed, the jet travels along the BH spin axis (perpendicular to the $x-y$ disk plane, i.e., along the $ z-$ direction). 
Figure~\ref{fig:angular_momentum}(c) shows the tilt and precession angles of both disk and one jet,  at $r=50\rg$ (dashed black line in Fig.~\ref{fig:angular_momentum}b). We calculate the disk tilt $\mathcal{T}_{\rm d}$ (orange line) and precession $\mathcal{P}_{\rm d}$ (orange circles) angles using $\mathcal{T}_{\rm d}={\cos^{-1}}  ( l_z / l )$ and $\mathcal{P}_{\rm d}={\cos^{-1}} [ l_y / (l_x^2+l_y^2)^{1/2} ]$, respectively. 
To compute the jet tilt and precession angles, we first find the Cartesian coordinates of the jet core centroid, at every radius $r$, using the magnetization-weighted average of $x_{\rm j},y_{\rm j},z_{\rm j}$, in regions where $\mu > 3$ and $b_{\rm p} \ge b_{\rm tor}$. 
The jet tilt is then given by $\mathcal{T}_{\rm j}={\cos^{-1}} ( z_{\rm j} / r_{\rm j} )$ (purple) and precession angle by $\mathcal{P}_{\rm j}={\cos^{-1}} [ y_{\rm j} / (x_{\rm j}^2+y_{\rm j}^2)^{1/2} ]$ (purple crosses), where $r_{\rm j}^2 = x_{\rm j}^2+y_{\rm j}^2+z_{\rm j}^2$.
Figure~\ref{fig:angular_momentum}(c) shows that during the MAD state (purple shaded) the tilt angle is small, $\mathcal{P_{\rm d}} \leq \pi/16$, at $t \lesssim 65$k: in other words, the angular momentum vector is nearly parallel to the \BH spin. For this reason, the disk precession angle, $\mathcal{P}_{\rm d}$, is ill-defined,  and we lower its opacity to minimize clutter. The jet tilt is, on average, similar to the disk tilt, with $\mathcal{T}_{\rm j} \leq \pi/16$. The jet precession angle shows high-amplitude variability but on average clusters around $\mathcal{P}_{\rm j} \simeq \pi/2$. This means that the jet on average has a preferred direction that is distinctly different than the rotational axis of the BH: indeed, Fig.~\ref{fig:angular_momentum} shows that both jets are skewed sideways, towards the $y$-axis.

At $50 \K \lesssim t \lesssim 65$k, the radial extent of the disk shrinks down to $\sim 10 \rg$, and by the end, it disappears, as the MAD state ends. This hardly seems like a coincidence, and we suggest that the system exits the MAD state when the accretion flow loses the rotational support. But what can cause the loss of angular momentum in the disk? It is possible that once launched, the jets perturb the infalling gas, and induce turbulence and vorticity, with angular momentum, on average, misaligned with the \BH spin vector. Once the misaligned angular momentum reaches the disk, it can cancel out the disk's coherent angular momentum, which was relatively low, to begin with, due to the limited disk radial extent ($R_{\rm disk} \simeq 50\rg \ll \rb = 1000\rg$). Moreover, strongly magnetized disks can launch winds that carry angular momentum outwards \citep{bp82} and can further reduce the disk's angular momentum. 
Whatever the case, the disk of small size, $\lesssim 10 \rg$, cannot support a jet with large-scale winds. The jet is left exposed to violent interactions with the ambient medium, for which the winds would normally act as a cushion. Additionally, the magnetic flux on the BH, once held by the disk, leaks out (Fig.\ref{fig:panel_fig}b) and the power of the jet decreases (Fig.\ref{fig:panel_fig}c).

From here on the system behavior changes drastically: at $t\gtrsim 65$k, our quasi-steady-state MAD transitions to a state dominated by the accretion of gas with a continuously varying direction of the angular momentum vector. For the duration, $69\K \lesssim t\lesssim 95$k, the disk tilt increases and the disk starts processing, while the angular momentum magnitude decreases. The flow with such low angular momentum is barely a disk, which we label as \statetwo state. The weakened and slower jet is now more affected by the presence of the ambient gas, and more specifically, the precessing and tilting disk start twisting the jet creating wiggles: this is similar to when one tags the end of a rope in a circular fashion creating circular waves that are transmitted along the rope. 
Around $t=78$k, the jets transiently fall apart (due to the shutoff of jet power at the \BH, Fig.~\ref{fig:panel_fig}c). Figure~\ref{fig:angular_momentum}(b) shows that subsequently, a short-lived disk of size $\sim 30\rg$ forms at $78 \K \lesssim t \lesssim 85$k. Figure~\ref{fig:panel_fig}(c) shows that the jets remain active until $t\simeq 95$k.

At $95 \K \le t \le 110$k, the disk gradually flips upside down and changes its sense of rotation from prograde ($\mathcal T_{\rm d}\sim 0$, rotating in the same sense as the \BH) to retrograde ($\mathcal T_{\rm d} \sim \pi$, rotating in the opposite sense to \BH). \footnote{Because the disk changes orientation, we show the tilt and precession of the counter jet.} This causes the jets to weaken, bend dramatically, and disrupt. We label this the rocking accretion disk state (\statethree), where the accretion of gas with strongly misoriented angular momentum is capable of flipping the disk upside-down. 
We highlight a specific part of the \statethree state, during  $110 \K \le t \le 130$k, where the disk gradually flips back to the prograde configuration, but stops at $\mathcal{T}_{\rm d}  \simeq \pi/2$. Interestingly, when looking at $r=15\rg$ (not shown here) the tilt reaches all the way to $\mathcal{T}_{\rm d}  \simeq \pi/4$. The absolute dimensionless magnetic flux, $\phi_{\rm BH}$, remains approximately constant during this transformation. In contrast, the magnetic flux through the northern hemisphere, $\phi_{\rm N}$, smoothly changes sign in phase with the disk tilt angle: this is precisely the behavior we would expect if the entire disk-jet system underwent a flip. During this stage of the simulation, we witness a large angular momentum inflow (Fig.~\ref{fig:angular_momentum}b), leading to an increase in the disk size to $\sim 200 \rg$. The formation of an accretion disk, which grows in size, deprives the BH of the gas and suppresses the BH accretion rate. 
Although $15$ times less powerful than in the MAD state, the weaker, intermittent and continuously reorienting jets fail to pierce through the infalling gas and are forced to bend around it, as seen in Fig.~\ref{fig:angular_momentum}(a3), they efficiently couple their energy to the infalling gas and additionally reduce the BH accretion. This results in BH mass accretion rate suppression to $\mdot/\mdotb \sim 1.7\%$, similar to the MAD state.

\section{External vs Internal Kink}
\label{sec:ext-vs-int-kink}

\begin{figure*}[!t]
\centering
\includegraphics[width=1\textwidth]{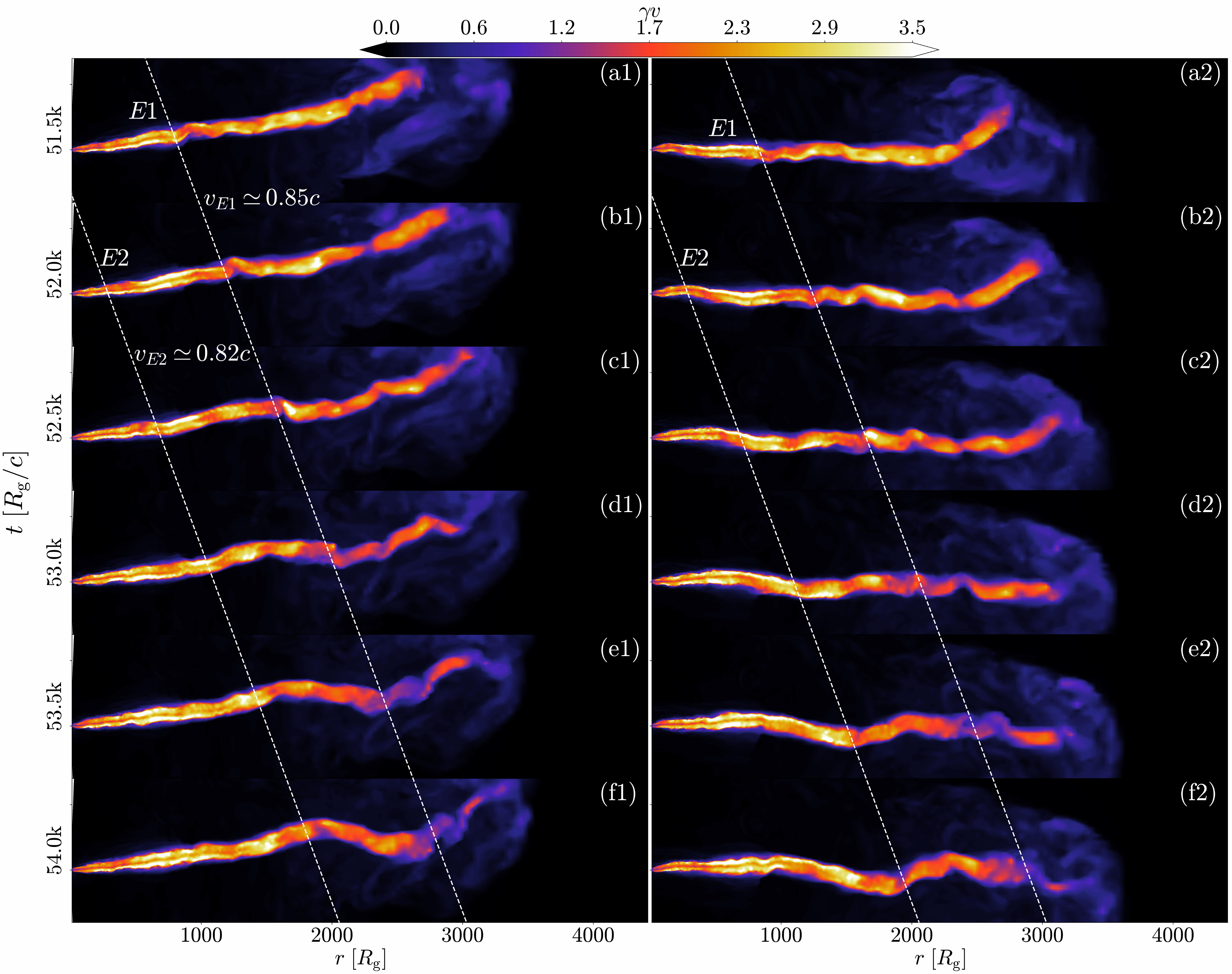}
\caption{Early-time MAD jet exhibits multiple small-amplitude bends and a major bend located at $r\simeq 1500\rg$. A movie is available in Supplementary Information and on \href{https://www.youtube.com/watch?v=8FXMQ5k044U&ab_channel=AretaiosLalakos}{YouTube (link)}.  The left and right columns show the jet's proper velocity, $\gamma v$, projected along the $y-, x-$ directions respectively. We fit a straight line through features along the jet, that show wiggles ($E1$, $E2$), and which we follow at different times to observe their growth. The slope of the white dashed line gives the propagation speed of the wobbles. For both $E1$ and $E2$ features, the velocity is approximately $v/c\simeq 0.82-0.85$. The wiggles, which can act as a seed for the kink instability, originate from the jet tilting and precessing about the BH spin. However, the fluid velocity is relativistic. $\gamma v \simeq 3$, and the jet-fluid does not have enough time to become kink-unstable, not until it reaches the distance of $r=1500\rg$, where it becomes globally-unstable.
}
\label{fig:spots_early}
\end{figure*}

\begin{figure*}[!t]
\centering
\includegraphics[width=1\textwidth]{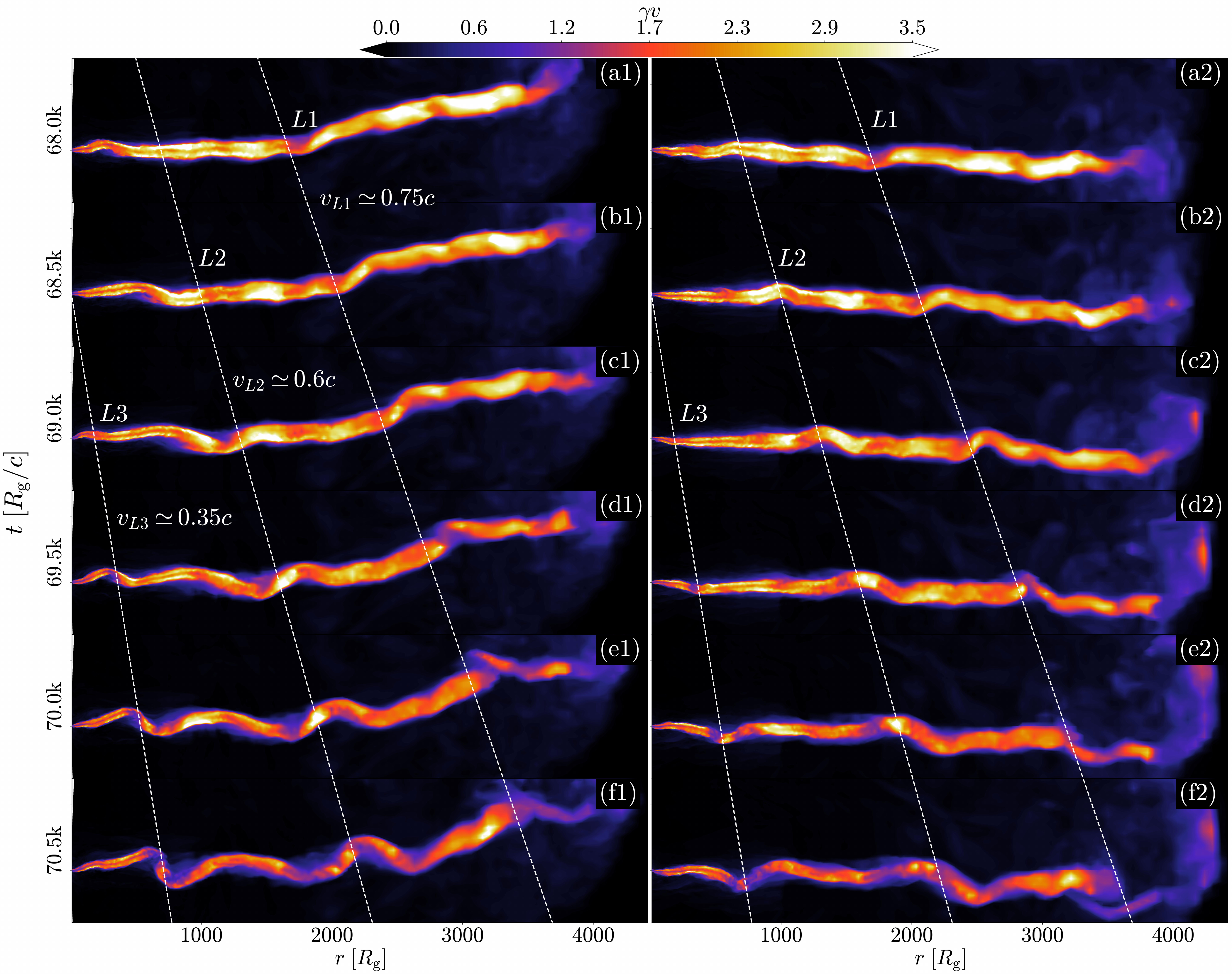}
\caption{Post-MAD jet gets progressively disrupted, with small bends fully developing into large bends.
Similar to Fig.~\ref{fig:spots_early}, the color shows the proper velocity, $\gamma v$, and the two columns are the projections along $y-$ and $x-$direction respectively. This jet has an average power of 4 times smaller than the average power during the MAD state. The drop in power, is also associated with the tilt of the disk slowly increasing, as shown in Fig.~\ref{fig:angular_momentum}(c). The power drop leads to jets that cannot efficiently accelerate to previously relativistic velocities, and as the jets follow the tilt of the disk, they run into the ambient medium, shock, and decelerate even more (see Fig.~\ref{fig:quant_pan}b2). External kink starts affecting the jet bends. As the bends get progressively larger, the fluid can become internally kink-unstable leading to the ``breaking'' of the jet. We fit a straight line through regions that develop large bends, $L1$, $L2$, and $L3$, and we estimate the wiggle propagation velocity from the slope. The $L1$ spot follows a bend developed during the end of the MAD state when the fluid was mostly stable to the kink. At $r\simeq 2000\rg$, a bend develops and the wiggle grows without significantly bending the jet until the jet fluid smashes at the head. The estimated propagation speed is slightly smaller than the MAD equivalent, at $v\simeq 0.75c$. The $L2$ spot propagates slower than $L1$, at $v\simeq 0.6c $ and the wiggle amplitude is noticeably larger. Finally, $L3$ is launched well into the weaker jet phase, and the jet bend becomes non-linear until the full jet disruption at $r\simeq 800 \rg$. The $L3$ wiggle propagates at $v\simeq 0.3c$, significantly slower than either $L1$ or $L2$, due to the fact that the large amplitude bends launched near the BH interacts with more ambient gas which slows it down even more.}
\label{fig:spots_late}
\end{figure*}

Figure~\ref{fig:spots_early} shows a time series of jet snapshots, illustrating the propagation of the jets and the growth of the kink instability during the early time in our simulation, in the MAD state ($31\K\le t\le 65$k): color shows the proper velocity, $\gamma v$, projected along the $y-$ and $x-$directions in left and right columns, respectively. 
We identify two early-time features, $E1$ and $E2$, that propagate along the jet and study how the kink instability grows in the jet. We pick these regions by either (i) visually searching for significant jet bends, or wiggles, or (ii) identifying jet regions with low values of the stability parameter, $\Lambda \lesssim 2$ (e.g., in a movie of Fig.~\ref{fig:quant_pan}b). 
In both projections, the jet exhibits multiple wiggles along its length, that originate near the jet launching region and propagate out. Such wiggles can emerge due to the stochastic nature of \BH accretion and wind-jet interactions (see, e.g., Fig.~\ref{fig:angular_momentum}c). Both $E1$ and $E2$ features originate as small-scale wiggles and propagate out to $r\simeq 1500\rg$. The jet remains mostly straight and maintains a relativistic velocity, $\langle \gamma v \rangle \simeq 3$. Such powerful jets as this one, in the MAD state, can accelerate to and sustain relativistic velocities ($\gamma v \gtrsim {\rm few}$): the relativistic motion stabilizes the jets against the kink instability via the relativistic time dilation (see eq.~\ref{eqn:growth_time0}, which has the Lorentz factor is in the numerator).
However, the jets propagate out to larger distances in a flat density distribution, they become more unstable (Sec.~\ref{sec:jet-stability-criterion}): after reaching $r\simeq 1500 \rg$, the jets develop unstable regions of $\Lambda \lesssim 2$ (Fig.~\ref{fig:quant_pan}), which implies that the kink instability growth timescale becomes comparable to the dynamical time of the jet, and the jets bends grow in amplitude. Indeed, both $E1$ and $E2$ features develop stronger jet bends that force the jet slows down, $\langle \gamma v \rangle \simeq 1.7$ (see eq.~\ref{eq:theta_bend}). These bends grow in amplitude and spatial scale, and culminate in essentially breaking up the jet into two segments, as seen in Fig.~\ref{fig:spots_early}(d) for feature $E1$ and in Fig.~\ref{fig:spots_early}(f) for feature $E2$.
To calculate the propagation speed of $E1$ and $E2$ features, we fit straight lines through them, as we show in Fig.~\ref{fig:spots_early}. Both features propagate along the jet at $v/c \simeq 0.8-0.85$, which is comparable but slightly lower than the jet velocity.


We have just discussed that in the MAD state the jets developed unstable features: jet bends that propagate outwards while slowly growing in amplitude. However, the jets managed to maintain their outward motion most of the way to the jet head. We also saw in Figs.~\ref{fig:kink_panels} and \ref{fig:quant_pan} that the jets globally disintegrated shortly after the system left the MAD state at $t\simeq 65$k. Can we develop a criterion to \emph{predict} when such catastrophic jet destruction is bound to happen? \citetalias{bromberg2016} derived a simple analytic approximation for a global jet stability parameter that evaluates the ability of the kink mode to grow on the jet periphery leading to a global deformation of the jet body. For a jet of power, $L_{\rm j}$, that moves through an ambient medium of constant density, $\rho_{\rm a}$, at a Lorentz factor, $\gamma=1/\sqrt{1-v^2/c^2}$, the stability parameter at the jet head distance, $r = r_{\rm h}$, is:
\begin{equation}
    \Lambda_{\rm h} \propto \left( \cfrac{L_{\rm j}}{\rho_{\rm a} r_{\rm h}^2 \gamma^2 c^3} \right) ^{1/6} \times \cfrac{\sqrt{v}}{v_{\rm A}} \,
    \label{eq:Lambda_lin}
\end{equation}
where for simplicity we have omitted constant prefactors.

Can eq.~\eqref{eq:Lambda_lin} predict the global jet disintegration we observe in our simulations? Upon exiting the MAD state, at $t=65$k, the jet power drops four-fold (Fig.~\ref{fig:panel_fig}c). At the same time,
the jet Lorentz factor decreases from $\langle \gamma v \rangle \simeq 3$ to $\langle \gamma v \rangle \simeq 1.7$, but $v_{\rm A}$ remains roughly unchanged (compare Figs.~\ref{fig:quant_pan}b1 and \ref{fig:quant_pan}b2). The distance to the jet head, $r_{\rm h}$ does not change significantly, because the drop in power occurs over a short timescale ($\Delta t\sim 5$k) and leaves the jet head little time to advance (compare Figs.~\ref{fig:kink_panels}b and \ref{fig:kink_panels}c). Thus, the stability parameter drops by an order unity factor $\Lambda_{\rm MAD}/\Lambda_{\rm tr} \simeq 1.11$,

Why does the stability criterion, eq.~\eqref{eq:Lambda_lin}, not capture the global jet instability we observe in the simulation? Figure~\ref{fig:spots_late} shows a time series of jet images at a later time, $68\K \le t \le 70.5$k, after the end of the MAD state and right up until the time when the jet becomes strongly kinked and globally unstable, a precursor to its global destruction. Figure~\ref{fig:spots_late}(a)--(c) shows a relativistic jet (yellow), $\langle \gamma v \rangle \simeq 3$, which eventually slows down to transrelativistic speed (orange),  $\langle \gamma v \rangle \simeq 1.7$, in Fig.~\ref{fig:spots_late}(d)--(f). The dark purple regions are locations where  $\langle \gamma v \rangle \lesssim 1$ and where the jet is getting disrupted. We identify three bend features ($L1$, $L2$, and $L3$) that grow in time. The feature $L1$ is one of the last features to be ejected during the MAD state. In Figs.~\ref{fig:spots_late}(a1,a2), the jet is mostly straight, apart from the large bend at $r\simeq 2000 \rg$: this is the bend associated with the $L1$ feature, and it grows as the feature approaches the jet head. The propagation speed of $L1$ is $v_{ L1}\simeq0.75c$, comparable to what we found for unstable regions at the earlier times, as seen in Fig.~\ref{fig:spots_early}. The $L2$ feature was ejected after the system exited the MAD state after the jets weakened and became unstable, enabling the $L2$ bend to grow, as seen in Fig.~\ref{fig:spots_late}(b2). Figure~\ref{fig:spots_late}(a)--(d) shows that as the $L2$ feature propagates out and grows in amplitude, its propagation speed decreases from that comparable to $L1$ down to $v_{L2}\simeq 0.6c$. Ejected at an even later time, the $L3$ feature starts out as a bend of high amplitude already at $r\lesssim500\rg$, in Fig.~\ref{fig:spots_late}(c). Its propagation speed, $v_{L3}\simeq 0.35c$, is the lowest among all $E$ and $L$ features and is, therefore, the most unstable. Indeed, Fig.~\ref{fig:spots_late} shows that the $L3$ feature grows to become a $90$-degree jet bend, at $r\simeq 800\rg$, that eventually globally disrupts the jet.

Summing up, we are witnessing a runaway growth of the kink instability that generates large-scale, strong jet bends. This implies that eq.~\eqref{eq:Lambda_lin}, for some reason does not apply to our weakened jets in the \statetwo state (red shaded region, $69\K\le t \le 95$k, in Fig.~\ref{fig:panel_fig}). Namely, once the system enters this state, the linear stability analysis (eq.~\ref{eq:Lambda_lin}) under-predicts the growth of the global kink. What is missing from the linear stability criterion, eq.~\eqref{eq:Lambda_lin}, are the non-linear effects that combine to cause a catastrophic run-away of the instability. Namely, the fluctuations in the disk tilt (Fig.~\ref{fig:angular_momentum}c) cause the jets to wobble around the \BH axis and propagate in directions offset from the jet pre-drilled path. This causes the jets to plow up the gas along new paths, through the shocked cocoon material. Because the cocoon was generated at earlier times when the jets were four times more powerful, the weakened jets have difficulty displacing the material, and get deflected: this causes the jets to bend and slow down (eq.~\ref{eq:theta_bend} and Fig.~\ref{fig:quant_pan}a1,b1). Slower velocity accelerates the kink (and bend) growth and further slows down jet propagation. As a result, such interactions lead to a runaway slowdown not only of the jet fluid but also of jet bends (Fig.~\ref{fig:spots_late}). In fact, the global kink instability we are witnessing here grows in the frame of the bends, which move slower than the jet fluid and hence more susceptible to the kink instability. 
This is an example of the non-linear development of an \emph{external} kink instability, which deforms the jet interface with the ambient medium. This is different than the \emph{internal} kink instability, which works on disrupting the jet core but leaves the jet-ambient medium interface intact. The two flavors of the kink instability can coexist, i.e., the development of the external kink instability can trigger the internal kink instability.

\begin{figure*}[!t]
\centering
\includegraphics[width=1\textwidth]{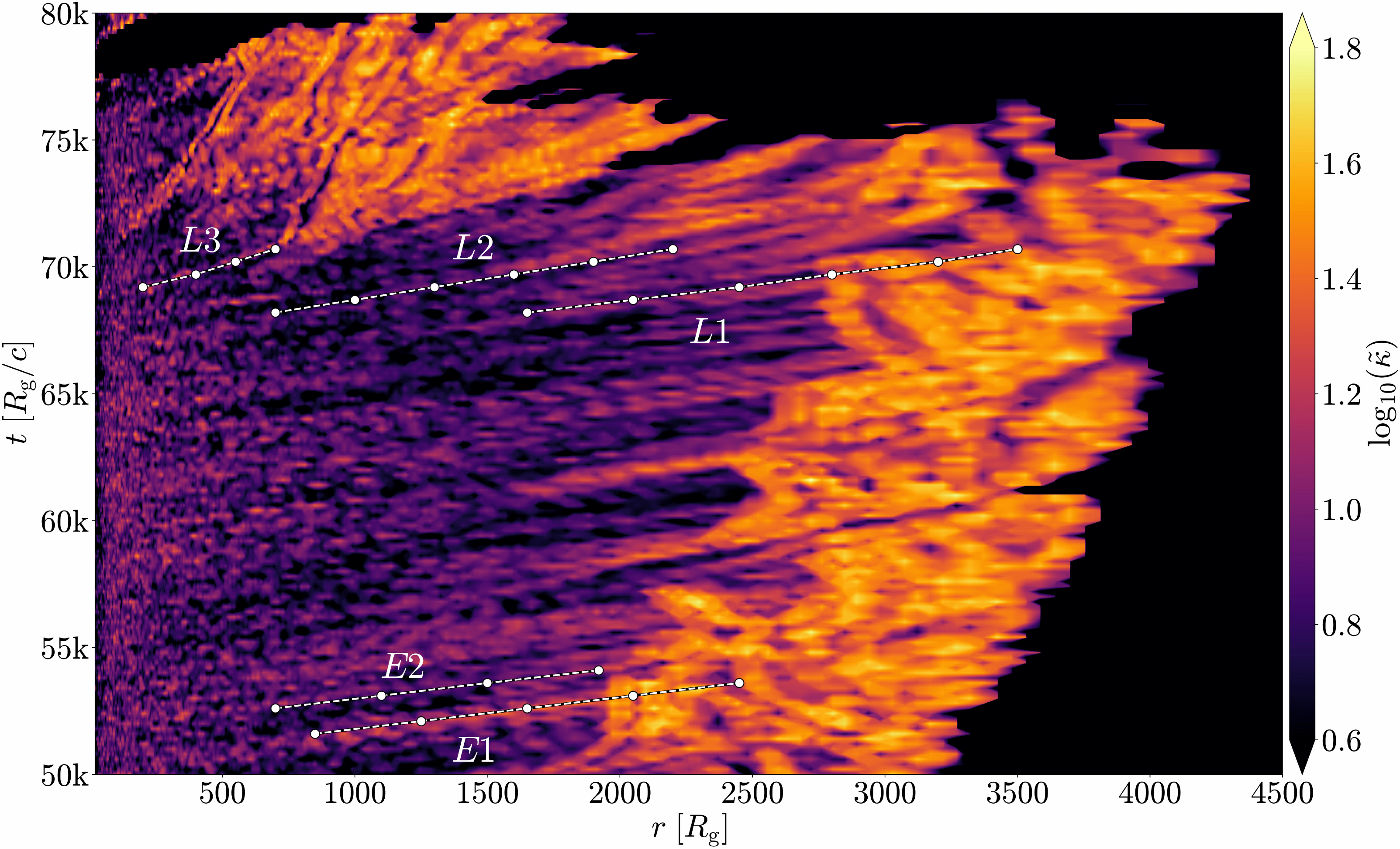}
\caption{Space-time diagram shows that normalized jet-curvature, $\Tilde{\kappa}$, grows due to the kink instability as it advects along the jet. We overplot the tracks of jet features in the MAD state ($E1$, $E2$) and \statetwo state ($L1$, $L2$, $L3$), also seen in Figs.~\ref{fig:spots_early}, \ref{fig:spots_late}. The slopes of these tracks give us the propagation speeds of the features. During the MAD state, the large-scale bends tend to develop at $r\simeq 1500\rg$, and propagate to larger scales at roughly the speed of the jet fluid, $v\sim 0.8c$, as the jet powers the cocoon. The cocoon outer boundary (right-most extent of the orange region boundary) expands at $v\simeq 0.05c$, i.e., much slower than the jet features. After the flow exits the MAD state ($t=65$k) and enters the \statetwo state ($t=69$k), the jet power drops, and strong jet bends form at much smaller distances, $r/\rg\simeq 500-1000$, and propagate at much slower speeds, $v\sim 0.3c$ (e.g., for feature $L3$). Eventually, the jet is globally destroyed, at $t\simeq 78$k, as its power vanishes (Fig.~\ref{fig:panel_fig}c).
}
\label{fig:spots_spacetime}
\end{figure*}

To more systematically study the nonlinear development of the global, external kink instability, Fig.~\ref{fig:spots_spacetime} presents a space-time diagram, where the color shows the logarithm of the dimensionless jet curvature,
%
\begin{equation}
    \tilde{\kappa} =  r \kappa = r \left| \cfrac{\der{ \vec{s}}}{dr}~ \right|,
\end{equation}
where $\kappa$ is jet curvature and $\vec{s}$ is a unit vector parallel to the jet. High values of the dimensionless curvature (yellow) indicate strong bends, whereas low values (black) correspond to mostly straight jets. We see that even at small distances, $r\lesssim 1000\rg$, the jet develops multiple bends along its length, due to the interactions with the ambient medium and disk winds.
We plot the tracks of all features we identified in Figs.~\ref{fig:spots_early}--\ref{fig:spots_late}: the slopes of the tracks give the bend propagation speeds. During the MAD state, $t\lesssim 65$k, the jet on average is mostly straight, and the bends travel at nearly the same speed as the jet fluid. If we follow the $E1$ and $E2$ features, they develop significant (orange) bends, at $r\simeq 1500 \rg$, where dimensionless curvature reaches $\log_{10}(\tilde\kappa)\gtrsim1.5$. 

Further away, at $r \gtrsim 2000\rg $, the jets get even more twisted: they become globally unstable, break apart, and energize the cocoon. Inflated by the jets, the cocoon expands at a non-relativistic speed of $v\simeq 0.05c$, as revealed by the slope of the jet-cocoon boundary, which is seen as the orange-black transition on the right of Fig.~\ref{fig:spots_spacetime}. In fact, Figure~\ref{fig:spots_spacetime} reveals that the entire bright orange region of the strongest jet bends shifts in time towards larger radii. This suggests that the proximity to the cocoon exacerbates jet bends: cocoon convective motions can displace and bend the jets sideways. Indeed, one can visually identify such convective cocoon motions in the movie of Fig.~\ref{fig:kink_panels}. Consistent with this picture, once the cocoon expands out to larger radii, the jets manage to avoid significant bends out to larger radii as well: the radius beyond which the jets start to show significant bends slowly increases from $r\simeq 1500\rg$ at $t=50$k to $r\simeq 3000 \rg$ at $t\simeq 68$k, just before the MAD state ends and the jets become globally unstable.

Visually following the slope of the ``late-time'' track of feature $L1$ in Fig.~\ref{fig:spots_spacetime}, we estimate the propagation speed of jet bends prior to disruption, $v_{L1} \simeq 0.8 c$, which is consistent with the best-fit values in Fig.~\ref{fig:spots_early}. 
However, once the MAD state comes to an end, at $t\simeq 65$k, the jet power drops. This causes each subsequent feature to travel at a slower speed, as indicated by the steepening of the slopes, decreasing to $v_{L2} \simeq 0.6c$ and eventually to $v_{L3} \simeq 0.35c$. This is consistent with the feature propagation speeds we measured in Fig.~\ref{fig:spots_late}. At $t\gtrsim70$k, the jet develops extreme bends and shortens to $r\lesssim2000\rg$, before globally disrupting at $t\simeq78$k. 

\section{Discussion and Conclusions} 
\label{sec:conclusions}

We study large-scale jet propagation and survival in global 3D GRMHD simulations of weakly magnetized, zero-angular momentum gas accretion onto rapidly-spinning BHs, $a=0.94$. The gas accretes from the Bondi-to-event horizon radius, traversing the largest scale separation to date, $\rb=10^3\rg$, in a single 3D GRMHD simulation.  Over time, the gas drags in from the ambient medium large-scale vertical magnetic flux, which readily accumulates on the BH and launches powerful relativistic collimated jets. 

Although initially our accretion flow started out without any angular momentum, as the accretion flow goes MAD, it forms an accretion disk of size $\sim 50\rg$ \citepalias[see also][]{ressler2021magnetically,2023ApJ...946L..42K}. Thus, no pre-existing accretion disk is required to form jets and collimate them via winds, as long as the BH is rapidly spinning and has accumulated dynamically-important large-scale poloidal magnetic flux on its event horizon. Interestingly, the MAD state achieved in our simulation is long-lived: its duration of $\Delta t=34\K$ is 3 times as long as the total simulation duration in \citetalias{ressler2021magnetically} and about half as long as the simulation duration in \citetalias{2023ApJ...946L..42K}, where in both cases the system only transiently enters the MAD state. 
Although our MAD state lasts comparatively longer, it still has a finite duration: the BH eventually loses the large-scale magnetic flux, and the jets fall apart. The MAD state ends in a fashion similar to that observed by \citetalias{ressler2021magnetically} and \citetalias{2023ApJ...946L..42K}. Once our jets transiently re-appear, they do not reach beyond the Bondi radius. 

To explore the origin of these differences, we repeated our simulation at a reduced scale separation, $\rb/\rg=10^2$, to match that of \citetalias[see Appendix~\ref{app:reduced_scale}]{ressler2021magnetically}.
We carried out the simulation for $t\simeq110\K$, a factor of $\sim 5$ longer than \citetalias{ressler2021magnetically}. The system turns MAD ($\phibh\gtrsim 50$) at $70\K \lesssim t \lesssim 100\K$, which suggests that the MAD state can establish itself, but it might take longer simulation durations in order to witness it. 
Why do MADs in the simulations of \citetalias{2023ApJ...946L..42K} survive for a rather short duration? One possibility is that our simulations contain more magnetic flux closer to the BH: although both simulations start with $\beta=100$ gas, we use a constant vertical field, $B_{z}={\rm constant}$, which results in the enclosed poloidal magnetic flux scaling as $\Phi \propto B_z r^2 \propto r^2$. In contrast, \citetalias{2023ApJ...946L..42K} use a parabolic-like field, $B_z \propto r^{-5/4}$, whose flux has a flatter radial profile, $\Phi\propto B_z r^2 \propto r^{3/4}$. Thus, out to the same distance, our simulations contain a larger flux reservoir. It is possible that this allows our simulations to transport the magnetic flux to the BH at a higher rate than the BH loses it due to the magnetic flux eruptions. Additionally, they have an adiabatic index $\Gamma=4/3$ describing radiation-dominated gas. Because this makes the gas more compressible, this can reduce the disk scale height and make it harder for the disk to hold on to the magnetic flux on the BH.

Apart from the early-time transient, our MAD jets carry the maximum power and extract \BH rotational energy at high efficiency, $\eta\simeq190\%$. 
The jets reach distances of $r\gtrsim 4000\rg$, or $3.5$ orders of magnitude in scale separation. In the inner $\sim50$\% of the jet length, $r \lesssim 1500\rg$, the jets remain mostly bend-free. However, further out, the jets develop twists and bends, as they become kink-unstable. Magnetic energy dissipates into thermal energy, and the jet magnetization drops until the magnetic fields reach equipartition with the gas thermal energy. Multiple locations along the jets show signs of recollimation and/or oblique shocks, associated with small-scale bends of the jets. To interpret this association, recall that at recollimation point the fluid has to decelerate, which can lead a marginally stable jet to become kink-unstable, and twist as a result. Figure~\ref{fig:quant_pan}(b) shows that eventually the instability grows non-linear and breaks up the jet into segments.  Because each segment is no longer constrained length-wise by the formerly adjacent jet segment, it is free to expand longitudinally and inject part of its energy into the cocoon, resulting in a dramatic drop in jet energy flux at the beginning of each segment.

The fact that the jets can reach $r\simeq 1500 \rg$ without getting globally disrupted is not trivial. \citetalias{ressler2021magnetically} find that their jets become kink unstable at $r\simeq 200-400\rg$. Using eq.~\eqref{eq:Lambda_lin} to extrapolate to larger Bondi radii and assuming a density profile of $\rho \propto r^{-1}$ within the Bondi radius, they predict that for $\rb \gtrsim 800\rg$,  jets should become kink-unstable. Because observations show many jets survive much greater distances, this would suggest that such jets cannot be powered by low-angular momentum accretion. Surprisingly, our jets make it out to $r\simeq 1500\rg$ without significant global bends. One possible reason for this difference is that when our jets first form, they navigate a steeper density profile, $\rho \propto r^{-3/2}$, than the one established at late times due to feedback processes, $\rho\propto r^{-1}$. The steeper profile makes it easier for the jets to escape unscathed \citepalias{bromberg2016}. Another difference is that soon after jet launching, the accretion flow goes MAD and maximizes the jet power. It is possible that, once our system exits the MAD state, and the jets transiently shut off, they would struggle to make it out to the same distances even if they went MAD again because the feedback causes the ambient density profile to flatten, $\rho \propto r^{-1}$, at late times. 

At $50\K \lesssim t \lesssim 65\K$, the MAD disk shrinks in size down to $\sim 10\rg$, the MAD state terminates ($\phibh < 50$), and the mass accretion rate increases. Initially, the jet power stays roughly constant, before dropping. Without the disk wind to shield the jets, the infalling gas slams into and bends them. The bends propagate out along the jets in a spiral pattern. Once the jet power drops, the mass accretion rate increases even further, and the system enters the \statetwo state at $69\K \lesssim t \lesssim 95\K$.  During this state, the jet becomes even more susceptible to bends, which are amplified by the external kink instability: in fact, the bends grow non-linear, drag against the cocoon material, and slow down. This leads to a run-away growth of the bends that twists the jets into a helix.
Thus, the low-angular momentum accretion system self-consistently forms precessing helical jets. We offer this as an observational alternative of periodical and curved jets, instead of binary companions or BH-torquing of a misaligned disk \citep{cui2023precessing}, especially since wobbles and precession can be hard to distinguish.

As the system went MAD, the jets reached the maximum energy efficiency and suppressed the accretion rate by a factor of $\sim70$: only a tiny fraction, $\mdot/\mdotb \simeq1.5\%$, of the Bondi accretion rate reaches the BH, and the feedback (e.g., by BH-powered jets and disk-powered winds) ejects the remaining $98.5\%$ of the gas.
\citetalias{Lalakos2022} used an identical setup, except for including non-zero angular momentum in the ambient medium (with circularization radius, $\rcirc=30\rg$), and found similar levels of $\mdot$ suppression. This suggests that most of the feedback is done by the powerful jets as opposed to disk winds. Indeed,
Fig.~\ref{fig:panel_fig}(a) shows that $\dot M$ remains roughly constant over the entire duration of the \statetwo state ($69\K\le t\le95\K$), despite the formation of an accretion disk at $78\K \lesssim t \lesssim 86\K$. In fact, we see that $\mdot$ transiently increases whenever the jet power vanishes, e.g., at $t\simeq70$k and $95\K\lesssim t\lesssim105\K$. This indicates that it is the jets, especially in the MAD state, that dominate the $\mdot$ suppression, as opposed to the rotationally supported disk and its wind.

At $t=71\K$, the power of the jet shuts off, and the jet gets globally destroyed. The subsequent jet formation has weak jets that mainly go through the pre-drilled path from the prior large-scale jets, with an average efficiency of $\eta \simeq 15\%$, until $t=94\K$. For the rest of the simulation, $t\gtrsim 95$k, the system enters the \statethree state, where the accreted material brings in gas with misaligned angular momentum, with respect to the BH spin. Newly formed intermittent and weaker jets don't surpass the Bondi radius intact, but promptly get disrupted, leaving behind weakly-magnetized buoyant bubbles. When the jets are not active, the mass accretion rate can reach all the way up to $\mdot /\mdotb \simeq 10\%$. Although the system cannot revert back to an apparent steady state, there are periods where the intermittent, and weaker, jets along with the formation of an ephemeral misaligned accretion disk, suppress the mass accretion rate down to values in the MAD state. For example, during $110\K \lesssim t \lesssim 130$k, the mass accretion rate averages to $\mdot/\mdotb = 1.7\%$, while the jet power is weaker compared to the MAD values by $\sim 15$. The outflow efficiency averages at $\eta =11\%$, while the normalized magnetic flux at $\phi_{\rm BH} =18$.

During the MAD state, the powerful and relativistic jets ($\gamma \simeq 4$) showcase extensive lobes and hotspots similar to FRIIs. During the \statethree state, the FRII jets have been disrupted at their base, and the new jets are intermittent, and propagate more slowly ($\gamma \simeq 2$) in varying directions without ever making it out of the Bondi radius ($r\lesssim 800\rg)$. This is supported by observations, with FR0 jets appearing mildly relativistic and with small sizes ($r\lesssim 1$kpc) making them hard to observe. For the first time, in a GRMHD simulation, we are witnessing a failed FRII jet transitioning into an FR0 state. Thus, we think that FR0 should be abundant in environments with extremely low angular momentum. Even though such systems are capable of creating FRI or FRII jets, if the disk loses its angular momentum and shrinks, the system inevitably turns \statethree. Additionally, the jets that are globally destroyed during the transition from \stateone to \statetwo, might be the source of $\gamma$-rays observed by a handful of FR0s \citep{Paliya_2021}, as a significant amount of the jet energy is dissipated into heat. The jets in the \statetwo and \statethree states might be significant contributors to the isotropic $\gamma-$ray background, the diffuse neutrino background (detectable by IceCube), and the ultra-high-energy cosmic rays. We will explore the radiation signatures of a globally kink-unstable jet, as well as the FR0 jets in a follow-up paper.

\section*{Acknowledgments}
AL wants to thank Nicholas Kaaz and Danat Issa for their fruitful ideas and discussions for this work.
OG is supported by the Flatiron Research and CIERA Fellowships.  OB was supported by an ISF grant 2067/22, a BSF grant 2018312, and an NSF-BSF grant 2020747 (OB \& AT). JJ and AT acknowledge support from the NSF AST-2009884 and NASA 80NSSC21K1746 grants. HZ was supported by NASA under award number 80GSFC21M0002.
AT was supported by 
NSF grants
AST-2107839,
AST-1815304,
AST-1911080,
AST-2206471,
OAC-2031997.
Support for this work was provided by the National Aeronautics and Space Administration through \chandra Award Number TM1-22005X issued by the \chandra X-ray Center, which is operated by the Smithsonian Astrophysical Observatory for and on behalf of the National Aeronautics Space Administration under contract NAS8-03060.
The authors acknowledge the Texas Advanced Computing Center (TACC) at
The University of Texas at Austin for providing HPC and visualization
resources that have contributed to the research results reported
within this paper via the LRAC allocation AST20011
(\url{http://www.tacc.utexas.edu}). 
This research used resources from the Oak Ridge Leadership Computing Facility, which is a DOE Office of Science User Facility supported under Contract DE-AC05-00OR22725. An award of computer time was provided by the ASCR Leadership Computing Challenge (ALCC), Innovative and Novel Computational Impact on Theory and Experiment (INCITE), and OLCF Director's Discretionary Allocation programs under award PHY129. This research used resources of the National Energy Research Scientific Computing Center, a DOE Office of Science User Facility supported by the Office of Science of the U.S. Department of Energy under Contract No. DE-AC02-05CH11231 using NERSC award ALCC-ERCAP0022634.

\appendix

\section{Fluid-frame magnetic fields}\label{App:comoving_fields}
Due to relativistic effects, such as aberration, we need to appropriately boost the lab-frame magnetic fields to compute the fluid-frame magnetic fields. We use the definitions: 
\begin{equation}
b^{t} = B^i u_i
\label{eqn:b_up_t}
\end{equation}
\begin{equation}
b^i = \cfrac{B^i+u^ib^{t}}{u^{t}}
\label{eqn:b_up_i}
\end{equation}
The 4-vector $b^{\rm \mu}$ is the fluid-frame magnetic field that is comoving with the 4-velocity $u^{\rm \mu}$. Since the linear analysis of the growth of the kink instability distinguishes the poloidal and toroidal components of the fields, we aim to do the same. In idealistic set-ups, the jet would mostly propagate along the z-axis, and one would write the magnetic field as $\vec{B}=\vec{B}_p+B_{\rm \varphi}\cdot \hat{e}_{\rm \varphi}$. Sufficiently far from the light cylinder, the bulk velocity of the jet would be in the poloidal direction. The Lorentz boost on the magnetic field at the lab frame to the comoving frame where a fluid element is moving with a Lorentz factor $\gamma_{\rm j}$ exclusively affects the toroidal component, while the poloidal remains unaffected; $b_{\rm \varphi}=B_{\rm \varphi}/\gamma_{\rm j}$ and  $b_{\rm p}=B_{\rm p}$. However, in our setup, the system is far from axisymmetry and the jets can be offset from the z-axis. Once developing wobbles and the jet axis deviates strongly from a straight line, simplistic assumptions on the morphology of the magnetic field prior to boosting in the fluid frame will lead to the wrong results. To address this hurdle, we first compute the drift velocity of the flow, and we will only include that for the Lorentz boost of the magnetic field component perpendicular to that velocity. We first define the base vectors $e^i_{\rm \|}$ and $e^i_{\rm \perp}$ that are parallel and perpendicular to the velocity of the flow respectively. That way we can write the 3-velocity as:
\begin{equation}
v^i = v_{\rm \|}^i+ v_{\rm \perp}^i =  v_{\rm \|} \cdot e^i_{\rm \|} + v_{\rm \perp} \cdot e^i_{\rm \perp}, 
\label{eqn:3-vel}
\end{equation}
where $e^i_{\rm \|} = B^i/B$;  $e^i_{\rm \perp}\cdot B^i = 0$; B is the magnitude of the lab-frame magnetic field $B^2 = B_i \cdot B^i$ with $B_i=g_{ij}B^j$. As mentioned above, the drift velocity will be perpendicular to the magnetic field component, hence subtracting the parallel component in eq.~(\ref{eqn:3-vel}) we get:  
\begin{equation}
v_{\rm dr} = v^i - v_{\rm \|}^i = v^i - \cfrac{(v^j \cdot B_j)B^i}{B^2},
\label{eqn:drift_vel}
\end{equation}
where $ v_{\rm \|}^i=(v^jB_j)B^i/B$ was obtained by multiplying eq.~\eqref{eqn:3-vel} with $B_i/B$.
After obtaining the spatial components of the drift velocity, we can find the time component of the proper velocity using the relativistic invariant $u_{\rm dr}^2 = (u^i\cdot u_i)_{\rm dr}=-1$: Expanding the product we get:
\begin{equation}
-1 = g_{tt}(u_{\rm dr}^{t})^2 + 2g_{{t}i} u_{\rm dr}^{t} u_{\rm dr}^i + g_{ii}(u_{\rm dr}^i)^2
\label{eqn:prod1}
\end{equation}
Noting that: $\gamma_{\rm dr}=u^{t}_{\rm dr}$ and $v_{\rm dr}^i = u_{\rm dr}^i/u_{\rm dr}^{t}$ we can rewrite eq.~(\ref{eqn:prod1}) as:
\begin{equation}
-1 = \gamma_{\rm dr}^2 \cdot \big ( g_{tt} + 2g_{{t}i}v_{\rm dr}^i + g_{ii}(v_{\rm dr}^i)^2 \big )
\label{eqn:prod2}
\end{equation}
Finally, solving for $\gamma_{\rm dr}$ in eq.~(\ref{eqn:prod2}), we calculate the fluid-frame magnetic fields as:
\begin{equation}
b_{\rm p} \simeq (B^r\cdot B_r)^{1/2}
\label{eqn:b_pol}
\end{equation}
\begin{equation}
b_{\rm \varphi} \simeq \dfrac{1}{\gamma_{\rm dr}}(B^2 - B^r\cdot B_r)^{1/2} 
\label{eqn:b_tor}
\end{equation}
Our approximation will hold true as long as 1) we are sufficiently far from the \BH $(r\ge 10\rg)$ so that most of the poloidal component is at the r-direction and not in the $\theta$-direction; 2) The jets have not developed extreme bends that result in the poloidal component having a significant contribution from the $\theta$ and $\varphi$ directions. We argue that in that case, if our goal is to study the growth of the kink instability in the linear regime, this case has ventured into the non-linear part where our equations should not hold.

\section{Adaptive Mesh Refinement Criterion}
\label{app:amr-criterion}

To identify the jet material, we use a cutoff in a proxy for entropy, $\Tilde{S}= u_{\rm g}/\rho^{\Gamma}$, where $u_{\rm g}$ is the internal energy density, $\rho$ is comoving frame density of the gas, and $\Gamma=5/3$ is the gas polytropic index for a monatomic non-relativistic gas. The reason is that magnetic dissipation in the jets increases the value of $\Tilde{S}$, making it appear as ``hot''.
For our run, the cut-off is chosen at $\Tilde{S}\ge 10$. We additionally refine the cocoons, created when the jets mix with the ambient gas resulting in a shocked and mildly-magnetized gas. The cut-off for the cocoons is chosen at $\Tilde{S}\ge0.1$. We also derefine blocks where the jets and/or cocoons left, when the entropy values dropped to $[50\%-100\%]$ of the cut-off. The AMR criterion is activated once the half-opening angle of either jets or cocoons becomes 
narrower than $48$ cells \citep[see][]{Gottlieb2022black2}, which is the width of a single AMR block, whose resolution is $N^{\rm B}_r\times N^{\rm B}_{\theta} \times N^{\rm B}_{\varphi}=56\times 48\times48$. The maximum effective resolution in the jets can reach $N^{\rm eff}_r\times N^{\rm eff}_{\theta} \times N^{\rm eff}_{\varphi}= 3,584\times768\times1,536$. This ensures that both the jets and their cocoons are well-resolved as they propagate out to large radii: jet opening angles of $0.04$~rad $=2.3^\circ$ are resolved with $\ge10$ cells out to distances of $5,000\rg$.

\section{Reduced scale separation}
\label{app:reduced_scale}
\begin{figure}[!ht]
\centering
\includegraphics[width=0.9\textwidth]{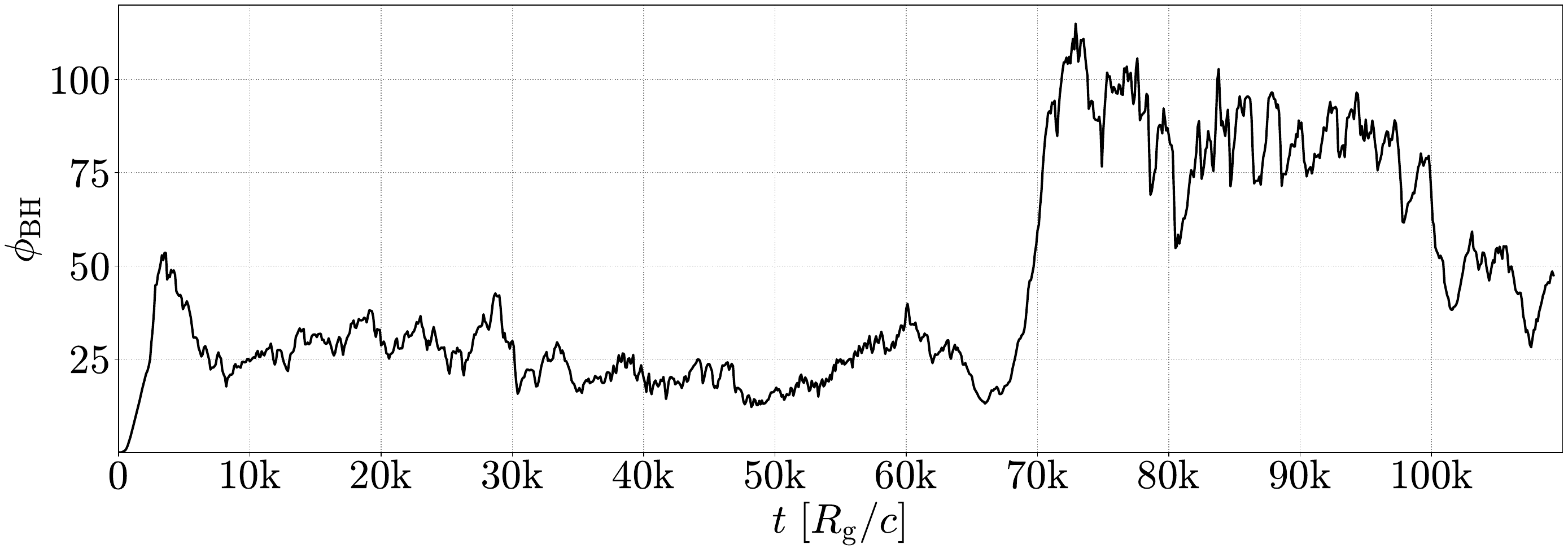}
\caption{For the reduced scaled separation, $\rb/\rg=10^2$, the system turns MAD at $70\K \lesssim t \lesssim 100\K$.}
\label{fig:panel_fig_100}
\end{figure}

We use the same physical parameters, as in the fiducial run, except we reduce the Bondi-to-event horizon scale separation by 1 order of magnitude, $\rb/\rg = 10^2$.
The grid setup is mostly identical to our fiducial run 
(see Sec.~\ref{sec:Numerical setup}). In this case, we did not tilt the BH spin $90^{\circ}$-degrees with respect to the polar axis. The base-grid resolution is the same: $N_r\times N_{\theta} \times N_{\varphi}=448\times96\times192$ cells in the \hbox{$r$-,} \hbox{$\theta$-,} and \hbox{$\varphi$-}directions. We activate 1 level of static mesh refinement (SMR) at $r\ge6.5\rg$, doubling the resolution in each dimension for an effective resolution $896\times192\times384$ cells.
Figure~\ref{fig:panel_fig_100} shows the normalized flux, $\phibh$, on the BH. The system turns MAD at $70\K \lesssim t \lesssim 100\K$.

\bibliography{mybib}{}
\bibliographystyle{aasjournal}



\end{document}